\documentclass[conference]{IEEEtran}
\IEEEoverridecommandlockouts
\usepackage{cite}
\usepackage{amsmath,amssymb,amsfonts}
\usepackage{algorithmic}
\usepackage{graphicx}
\usepackage{textcomp}
\usepackage{xcolor}
\usepackage[graphicx]{realboxes}
\usepackage[figuresleft]{rotating}
\usepackage{balance}
\usepackage{url}
\usepackage{tikz}
\usepackage{multirow}
\usetikzlibrary{positioning, fit, arrows.meta, calc, shapes.geometric, backgrounds} 
\usepackage[normalem]{ulem}
\usepackage{booktabs}
\usepackage{svg}
\usepackage[caption=false,font=footnotesize]{subfig}
\usepackage{circuitikz}
\usepackage{etoolbox}
\usepackage{adjustbox}

\def\BibTeX{{\rm B\kern-.05em{\sc i\kern-.025em b}\kern-.08em
    T\kern-.1667em\lower.7ex\hbox{E}\kern-.125emX}}

\usepackage[bookmarks=true,hidelinks,pdfpagelabels=true]{hyperref}
\usepackage[printonlyused,nohyperlinks]{acronym}

\begin{document}
\bstctlcite{BSTcontrol}

\title{
%Experimental Demonstration of 5G Sensing for Human Pose Estimation 
5G Sensing for Human Pose Estimation with PRS-based CSI: An Experimental Demonstration
\thanks{
This work is supported by the MSCA TMA DN SMARTTEST project funded from the European Union’s Horizon Europe - Research and Innovation program - under grant agreement no. 101167834.
}
}
\author{\IEEEauthorblockN{Zhaorui Yin, Pietro Morri, Mattia Brambilla, Monica Nicoli}
\IEEEauthorblockA{Politecnico di Milano, Milan, Italy}}

% \author{
% \IEEEauthorblockN{Zhaorui Yin}
% \IEEEauthorblockA{
%     % Department of XXX\\
%     Politecnico di Milano\\
%     Milan, Italy\\
%     zhaorui.yin@polimi.it
% }
% \and
% \IEEEauthorblockN{Mattia Brambilla}
% \IEEEauthorblockA{
%     % Department of XXX\\
%     Politecnico di Milano\\
%     Milan, Italy\\
%     mattia.brambilla@polimi.it
% }
% \and
% \IEEEauthorblockN{Monica Nicoli}
% \IEEEauthorblockA{
%     % Department of XXX\\
%     Politecnico di Milano\\
%     Milan, Italy\\
%     monica.nicoli@polimi.it
% }
% }

\maketitle

\acrodef{AWGN}[AWGN]{Additive White Gaussian Noise}
\acrodef{BLL}[BLL]{Bone Length Loss}
\acrodef{CFR}[CFR]{Channel Frequency Response}
\acrodef{CNN}[CNN]{Convolutional Neural Network}
\acrodef{CP}[CP]{cyclic prefix}
\acrodef{coco17}[COCO17]{Common Objects in Context 17}
\acrodef{CSI}[CSI]{Channel State Information}
\acrodef{CIR}[CIR]{Channel Impulse Response}
\acrodef{DFT}[DFT]{Discrete Fourier Transform}
\acrodef{DL}[DL]{Deep Learning}
\acrodef{DFS}[DFS]{Doppler Frequency Spectrum}
\acrodef{FFT}[FFT]{Fast Fourier Transform}
\acrodef{IDFT}[IDFT]{Inverse DFT}
\acrodef{FMCW}[FMCW]{Frequency Modulated Continuous Wave}
\acrodef{ofdm}[OFDM]{Orthogonal Frequency-Division Multiplexing}
\acrodef{ISAC}[ISAC]{Integrated Sensing and Communication}
\acrodef{LS}[LS]{least-squares}
\acrodef{GT}[GT]{Ground Truth}
\acrodef{GCN}[GCN]{Graph Convolutional Network}
\acrodef{MAE}[MAE]{Masked Autoencoder}
\acrodef{MHSA}[MHSA]{Multi-Head Self-Attention}
\acrodef{MIMO}[MIMO]{Multiple-Input Multiple-Output}
\acrodef{MUSIC}[MUSIC]{Multiple Signal Classification}
\acrodef{MLP}[MLP]{Multilayer Perceptron}
\acrodef{MPJPE}[MPJPE]{Mean Per-Joint Position Error}
\acrodef{MSE}[MSE]{Mean Squared Error}
\acrodef{PA-}[PA-]{Procrustes-aligned}
\acrodef{PDP}[PDP]{Power Delay Profile}
\acrodef{PCK}[PCK]{Percentage of Correct Keypoints}
\acrodef{PA-MPJPE}[PA-MPJPE]{Procrustes-aligned MPJPE}
\acrodef{SOTA}[SOTA]{state-of-the-art}
\acrodef{USRP}[USRP]{Universal Software Radio Peripheral}
\acrodef{sg}[SG]{Savitzky--Golay}
\acrodef{ViT}[ViT]{Vision-Transformer}
\acrodef{lpr}[LPR]{Linear Phase Removal}

%%%%%%%%%%%% 5G %%%%%%%%%%%%%%
\acrodef{5g}[5G]{Fifth Generation}
\acrodef{ul}[UL]{Uplink}
%\acrodef{dl}[DL]{Downlink}
\acrodef{gnb}[gNB]{Next Generation Node Base}
\acrodef{grc}[GRC]{GNRadio Companion}
\acrodef{pbch}[PBCH]{Physical Broadcast Channel}
\acrodef{prb}[PRB]{Physical Resource Block}
\acrodef{prs}[PRS]{Positioning Reference Signal}
\acrodef{rb}[RB]{Resource Block}
\acrodef{scs}[SCS]{Subcarrier Spacing}
\acrodef{srs}[SRS]{Sounding Reference Signal}
\acrodef{SSB}[SSB]{Synchronization Signal Block}
\acrodef{ue}[UE]{User Equipment}
\acrodef{SDR}[SDR]{Software-Defined Radio}

% \begin{abstract}
% We study \ac{5g} \ac{prs}-based \ac{CSI} for device-free 2D human pose estimation. The \acp{gnb} transmit downlink \ac{prs} signals, while one \ac{ue} with two receive antennas collects the received signal. The \ac{CSI} is first estimated by a \ac{LS} method. Then, a \ac{CSI}-ratio is computed between the two receive antennas to reduce common hardware phase offsets and correlated amplitude disturbances. The resulting amplitude and phase tensors are used as input to two learning-based pose models, MetaFi++ and DT-Pose, under a unified COCO17 2D keypoint output. We evaluate amplitude-only and joint amplitude-phase inputs with and without \ac{CSI}-ratio processing for bandwidths of 20, 40, and 60 MHz. Validation is reported using \ac{PCK}, \ac{MPJPE}, \ac{PA-MPJPE}, and \ac{BLL}. TODO results.
% \end{abstract}

\begin{abstract}
%\textcolor{orange}{To Mattia: what does "bstctlcite\{BSTcontrol\}" do? And corrected abstract according to Prof. Nicoli`s suggestions}
This paper presents an experimental study of device-free 2D human pose estimation using \ac{5g} \acp{prs}. Human pose is inferred from the \ac{CSI} of the radio link, without requiring the subject to carry or wear any device.
%The proposed system employs a \ac{gnb} transmitting \acp{prs} to a \ac{ue} equipped with two receive antennas to measure the radio channel, and passively estimates the pose of a human in the environment. 
%A camera simultaneously provides ground-truth human poses in the COCO17 format. 
%The \ac{CSI} is estimated using a \ac{LS} method, %aligned with the camera frames, 
%and processed through amplitude smoothing and phase calibration. 
The \ac{CSI} is estimated using a \ac{LS} method, and various amplitude and phase pre-processing configurations are evaluated. The resulting \ac{CSI} representations are then used as inputs to a deep learning model for pose estimation.
% The resulting amplitude and phase representations are used as inputs to a deep learning model for pose estimation.
We compare amplitude-only and joint amplitude-phase inputs using a bandwidth of 20\,MHz.
%and provide a preliminary evaluation at 40\,MHz. At 20\,MHz, 
%The results show that dedicated pre-processing are not required for the considered setup. 
The best performance is obtained without the pre-processing steps, with the joint phase-amplitude approach outperforming
the amplitude-only solution. Specifically, it achieves \ac{PCK} values of 51.5\%, 83.2\%, and 91.2\% for thresholds corresponding to 10\%, 30\%, and 50\% of the distance between the left shoulder and the right hip, respectively. 
% We show that the joint use of phase and amplitude outperforms amplitude-only solution. 
% \textcolor{blue}{The full 40\,MHz improves amplitude-only performance, whereas joint training shows a gain only at \ac{PCK}@10 and lower accuracy at \ac{PCK}@30 and \ac{PCK}@50.} 
%The per-keypoint analysis shows higher accuracy for the torso and lower-body joints, while the wrists and elbows remain more difficult to estimate. 
These results demonstrate that the \ac{CSI} derived from the standardized downlink \ac{5g} \ac{prs} provides useful information for human pose estimation, even with a simple radio configuration and a relatively small dataset.
\end{abstract}

\acresetall

\begin{IEEEkeywords}
5G sensing, CSI, PRS, Pose estimation, Deep Learning
\end{IEEEkeywords}

\section{Introduction}
\label{sec:intro}

\ac{5g} cellular systems are progressively evolving beyond their traditional role as communication networks to support positioning, sensing, and environmental awareness~\cite{JPCAS}. 
Standardized signals and positioning procedures provide a foundation for such capabilities~\cite{3gpp38211}, and recent research has highlighted the potential of leveraging physical-layer measurements of \ac{5g} networks for positioning  \cite{Brambilla_GNSS} and sensing applications~\cite{5g-prs-sensing1, 5g-prs-sensing2, midipose}. These developments align with the \ac{ISAC} paradigm, where the same wireless infrastructure supports connectivity, localization, and environmental perception.

Among \ac{5g} signals, the \acp{prs} play a central role in enabling network-based localization. They are transmitted by the \acp{gnb} and allow the receiver to estimate the \ac{gnb}-to-\ac{ue} channel; their transmissions are specifically designed to facilitate time-of-arrival estimates through configurable resource allocation and flexible bandwidth configurations~\cite{LorenzoBest}.  As a result, \ac{prs} measurements contain detailed information about the propagation environment and the objects interacting with the radio channel. While these signals have been extensively studied for positioning purposes, their suitability for fine-grained human sensing remains largely unexplored.

Recent advances in wireless sensing have demonstrated that radio-frequency channel measurements can capture information about human motion and activities~\cite{savazzi2013bayesian, savazziDeviceFreeRadioVision2016,guarino2026survey}. In particular, deep learning approaches have shown that \ac{CSI} can be directly mapped to human skeletal representations, enabling device-free pose estimation and activity recognition without requiring cameras or wearable devices~\cite{personinwifi, geng2022densepose, zhou2023metafi++, chenRobustRealisticHuman2025, yin2026, diaz2023phase}. Most existing approaches, however, have been developed using Wi-Fi sensing platforms and datasets specifically designed for wireless perception. Consequently, their applicability to cellular systems operating under standardized \ac{5g} signaling remains an open research question.

The transition from Wi-Fi-based sensing to \ac{5g} sensing is not straightforward. In \ac{5g}, \ac{CSI} is extracted from structured pilot resources within a network-controlled frame structure. The pilot allocation, bandwidth, transmitter geometry, and synchronization conditions are different from typical Wi-Fi sensing systems. Understanding whether \ac{CSI} extracted from standardized \ac{5g} \acp{prs} contains sufficient information for recovering human body pose is therefore an important step toward network-native sensing services. MiDiPose~\cite{midipose} represents a recent effort in this direction, using uplink \ac{srs}-based \ac{5g} measurements for 2D human pose estimation. It extracts amplitude, phase, and Doppler-domain features from \ac{CSI}, aligns these engineered feature vectors in a shared latent space, and fuses them for keypoint regression. However, downlink \ac{prs}-based \ac{CSI} for 2D human pose estimation is still not well studied.

% \textcolor{orange}{TODO: What MiDiPose did, what I could do and why I do this and how it differs: we do not use vector but image based input and ViT.}

%Our work is complementary to MiDiPose in the radio setting and sample number. 
Instead of measurements collected from a multi-receiver configuration, we consider a setup consisting of a single-antenna \ac{gnb} and a two-antenna \ac{ue}.
Although this configuration provides less spatial diversity, it allows us to investigate whether a single standardized \ac{prs} downlink already carries sufficient information for pose discrimination. While MiDiPose uses more than $200,000$ samples in our case, we evaluate the proposed approach with a much lower sample number, to see whether  accurate pose estimation can still be achieved with a short sensing time as typically required for activities performed in daily life. 
% In contrast to the larger MiDiPose dataset, we evaluate feasibility using a substantially smaller dataset, to see whether accurate pose estimation can still be achieved with a short sensing time as typically required for activities performed in daily life. 

% \textcolor{blue}{
% Another open question is the role of bandwidth. In Wi-Fi sensing, reducing the bandwidth does not always degrade \ac{CSI}-based human activity recognition performance~\cite{cominelli2023exposing}. For \ac{5g}, this question is also important because different bandwidths can be allocated depending on spectrum availability and service requirements, while wider bandwidths provide finer delay resolution~\cite{resource_alloc}. It is therefore relevant to understand whether pose-estimation accuracy benefits from additional bandwidth or whether reliable sensing can already be achieved using narrower allocations. Such an analysis is particularly important for future \ac{ISAC} deployments. 
% %where sensing and communication functionalities must coexist within the same radio resources. 
% }

In this work, we investigate human pose estimation from \ac{CSI} extracted from downlink \ac{5g} \ac{prs}. A bistatic experimental platform based on \acp{SDR} is employed to collect \ac{CSI} measurements, to be used as input to a \ac{DL} pose-estimation model originally developed for Wi-Fi sensing.
The achieved results corroborate the experimental feasibility of recognizing human skeleton with \ac{5g} signals from standardized downlink \ac{prs}-based \ac{CSI} in a configuration with a single \ac{gnb} and two receive antennas, 
unlocking a new  application area of the  cellular wireless technology.

\section{\ac{5g} Sensing System and Dataset}

\subsection{Radio Sensing Scenario}

We consider a \ac{5g} radio sensing system for human pose estimation, where \acp{prs} are transmitted in downlink by one \ac{gnb} and received by a \ac{ue} equipped with two antennas. 
The scenario includes a human performing different poses in the sensing area.
The received signals are first converted into \ac{CSI} features, with amplitude and phase branches later used by the pose-estimation algorithm.
At the same time, a camera supervises the scene, and outputs human skeleton keypoints using the YOLO26 algorithm \cite{yolo26}.

The transmitter \ac{gnb} is scheduled to transmit a \ac{prs} within one slot of a \ac{5g} frame. The \ac{prs} configuration uses a comb 
%\sout{size} 
spacing
of $K_{\mathrm{COMB}}^{\mathrm{PRS}}=2$ and $L_{\mathrm{PRS}}=3$, 
where $L_{\mathrm{PRS}}$
%\sout{$=3$} 
denotes the number of \ac{ofdm} symbols carrying \ac{prs} within the scheduled slot. 
This pattern is illustrated in Fig.~\ref{fig:5g_prs}, where the second slot is shown as an example. In the slot, three \ac{ofdm} symbols are allocated for \ac{prs} pilots, represented by the green resource elements. The \ac{prs} pilots are modulated and transmitted over the air.
The grey resource elements are not used for \ac{prs} in the current setup. 
They can be reserved for future \ac{ISAC} extensions, for example by carrying \ac{5g} data symbols, or by supporting a future multistatic configuration with additional transmitting \acp{gnb}.

% \textcolor{orange}{the following paragraph is to be removed due to one tx one rx (2 antennas at the rx) setup. Backup file is stored under folder "old"}
% \sout{
%Note that in the first slot, the first four \ac{ofdm} symbols are reserved to \ac{pbch} for synchronization, and the remaining symbols are used for \ac{prs}, where symbols $5$--$8$ are assigned to \ac{gnb}~1 and symbols $9$--$12$ are assigned to \ac{gnb}~2. For slots $2$ to $20$, symbols $1$--$4$ are assigned to \ac{gnb}~$1$, symbols $5$--$8$ are assigned to \ac{gnb}~$2$, and symbols $9$--$12$ are assigned to \ac{gnb}~$3$. Symbols $13$--$14$ are not used for the considered \ac{prs}-based \ac{CSI} extraction. This symbol-level time-division schedule avoids inter-\ac{gnb} \ac{prs} collision and allows the receiver to estimate the \ac{dl} channel of each \ac{gnb} separately. The missing data in slot $1$ for \ac{gnb}~$3$ is duplicated with the data in slot $2$.} \textcolor{red}{what do you mean it is duplicated?}
% \textcolor{orange}{For every first slot, 14 symbols, 4 for SSB, 4 for gNB1 prs, 4 for gNB2 prs, and nothing left for the gNB3 so I thought about copying the value of gNB3 in slot 2 to slot 1. }
% At sensing frame $t=1,\ldots,N_{\mathrm{t}}$, after \ac{ofdm} demodulation, the received resource-grid sample at receive antenna $r\in\{1,2\}$, \ac{ofdm} symbol $l$, and subcarrier $k=1,\ldots,N_{\mathrm{K}}$ is denoted by $Y_{r,t}[k,l]$. 

For each demodulated \ac{ofdm} pilot symbol, the received resource-grid sample at \ac{ue} receive antenna $a\in\{1,2\}$, \ac{ofdm}-symbol index $l\in\{1,\ldots,14\}$ , and subcarrier $k=1,\ldots,N_{\mathrm{K}}$ is denoted by $Y_{a}[k,l]$. 
Numerology $\mu=1$ is employed, corresponding to a \ac{scs} of $30$~kHz, and $20$ slots in a \ac{5g} frame of $10$ ms~\cite{3gpp38211}.
The \ac{prs} bandwidth of $B = 20~$MHz is considered. Its expression in terms of \acp{rb} is:  $N_{\mathrm{PRB}} = 51$ \acp{prb}. Therefore, the number of usable subcarriers is
$N_{\mathrm{K}} = 12\times N_{\mathrm{PRB}}  = 612$.
% The following \ac{prs} bandwidths are considered: $B\in\{20,40\}~$MHz. Their expression in terms of \acp{rb} is:  $N_{\mathrm{PRB}}\in\{51,106\}$ \acp{prb}. Therefore, the number of usable subcarriers is $N_{\mathrm{K}} = 12\times N_{\mathrm{PRB}} \in \{612,1272\}$.
% After averaging the \ac{prs} pilot symbols within each slot, the resulting slot-level \ac{CSI} sample is indexed by $s\in\{1,\ldots,N_{\mathrm{s}}\}$.
%Since one slot lasts only $0.5~\mathrm{ms}$ for numerology $\mu=1$, the human pose is assumed to be approximately unchanged within this short time. 
% \textcolor{orange}{TODO: to change to only one gNB so the following paragraph to be modified}
The received samples are used for a \ac{LS} channel estimation, and averaged during pre-processing per slot, then per frame.

% The overall pipeline is shown in Fig.~\ref{fig:csi-dtpose}.

\usetikzlibrary{calc,arrows.meta,decorations.pathreplacing,positioning,matrix}

\usetikzlibrary{backgrounds}

%\definecolor{usrpblue}{RGB}{64,113,175}
%\definecolor{usrppurple}{RGB}{124,95,158}
%\definecolor{usrporange}{RGB}{239,137,68}
\definecolor{usrpblue}{rgb}{0.36, 0.54, 0.66}
\definecolor{usrppurple}{rgb}{0.9, 0.17, 0.31}
\definecolor{usrpgreen}{rgb}{0.0, 0.5, 0.0}

\begin{figure}[t]
    \centering
    \resizebox{0.99\linewidth}{!}{%
\begin{tikzpicture}[font=\small, >=Latex]

\newcommand{\txantennaLeft}[3]{%
  \begin{scope}[shift={(#1,#2)},color=#3,rotate=180]
    \draw[very thick] (0,0) -- (0,0.6);
    \draw[very thick] (0,0.6) -- (0.2,0.9);
    \draw[very thick] (0,0.6) -- (-0.2,0.9);
    \draw[very thick] (0.3,0.5) arc[start angle=-30,end angle=25,radius=0.4];
    \draw[very thick] (0.4,0.4) arc[start angle=-30,end angle=25,radius=0.6];
  \end{scope}%
}

\newcommand{\txantenna}[3]{%
  \begin{scope}[shift={(#1,#2)},color=#3]
    \draw[very thick] (0,0) -- (0,0.6);
    \draw[very thick] (0,0.6) -- (0.2,0.9);
    \draw[very thick] (0,0.6) -- (-0.2,0.9);
  \end{scope}%
}

% ============================================================
% Left: radio frame + slot-2 PRS allocation
% ============================================================

% --- top radio frame ---
\def\fx{0.5}
\def\fy{8.58}
\def\sw{0.44}
\def\sh{0.46}

\txantennaLeft{\fx+20*\sw}{\fy+\sh-0.3}{usrpblue}

\node[anchor=west] at (\fx,9.36) {\ac{5g} frame ($10$ ms)};

\draw[thick] (\fx,\fy) rectangle ({\fx+20*\sw},{\fy+\sh});
\foreach \i in {0,...,19} {
  \pgfmathtruncatemacro{\slotnum}{\i+1}
  \ifnum\slotnum=2
    \fill[usrpblue!30] ({\fx+\i*\sw},\fy) rectangle ({\fx+(\i+1)*\sw},{\fy+\sh});
    \draw[very thick, usrpgreen!75] ({\fx+\i*\sw},\fy) rectangle ({\fx+(\i+1)*\sw},{\fy+\sh});
  \else
    \draw[thin, black!70] ({\fx+\i*\sw},\fy) rectangle ({\fx+(\i+1)*\sw},{\fy+\sh});
    \fill[usrpblue!30] ({\fx+\i*\sw},\fy) rectangle ({\fx+(\i+1)*\sw},{\fy+\sh});
  \fi
}

\txantenna{\fx+20*\sw-1.5}{\fy+\sh-2.5}{usrpblue}
\txantenna{\fx+20*\sw-0.2}{\fy+\sh-2.5}{usrpblue}

\foreach \i/\lab in {0/1,1/2,2/3,3/4,4/5,9/10,14/15,19/20} {
  \node[font=\scriptsize] at ({\fx+(\i+0.5)*\sw},{\fy+0.5*\sh}) {\lab};
}

% 2nd grid highlight + arrow
% \draw[->, thick, red!75] ({\fx+1.5*\sw},{\fy-0.04}) -- ({\fx+1.5*\sw},8.00);

% --- 2nd grid moved upward ---
\def\x0{0.5}
\def\y0{3}
\def\cw{0.44}
\def\ch{0.37}
\def\cols{14}
\def\rows{12}
\pgfmathsetmacro{\gridtop}{\y0+\rows*\ch}

\fill[usrpblue!10] (\x0,\y0) rectangle ({\x0+\cols*\cw},{\y0+\rows*\ch});
\draw[usrpgreen, line width=1pt] (\x0,\y0) rectangle ({\x0+\cols*\cw},{\y0+\rows*\ch});

% gNB1 symbols 1..4
\foreach \c in {0,1,2} {
  \pgfmathtruncatemacro{\offset}{mod(\c,2)}
  \foreach \r in {0,1,...,11} {
    \pgfmathtruncatemacro{\active}{mod(\r+\offset,2)}
    \ifnum\active=0
      \fill[usrpgreen!40] ({\x0+\c*\cw},{\y0+\r*\ch}) rectangle ({\x0+(\c+1)*\cw},{\y0+(\r+1)*\ch});
    \fi
  }
}
% gNB2 symbols 5..8
% \foreach \c in {4,5,6,7} {
%   \pgfmathtruncatemacro{\offset}{mod(\c-4,2)}
%   \foreach \r in {0,1,...,11} {
%     \pgfmathtruncatemacro{\active}{mod(\r+\offset,2)}
%     \ifnum\active=0
%       \fill[usrppurple!85] ({\x0+\c*\cw},{\y0+\r*\ch}) rectangle ({\x0+(\c+1)*\cw},{\y0+(\r+1)*\ch});
%     \fi
%   }
% }

% gNB3 symbols 9..12
% \foreach \c in {8,9,10,11} {
%   \pgfmathtruncatemacro{\offset}{mod(\c-8,2)}
%   \foreach \r in {0,1,...,11} {
%     \pgfmathtruncatemacro{\active}{mod(\r+\offset,2)}
%     \ifnum\active=0
%       \fill[usrpgreen!88] ({\x0+\c*\cw},{\y0+\r*\ch}) rectangle ({\x0+(\c+1)*\cw},{\y0+(\r+1)*\ch});
%     \fi
%   }
% }

% grid lines
\foreach \i in {1,...,13} {
  \draw[thin, gray!60] ({\x0+\i*\cw},\y0) -- ({\x0+\i*\cw},{\y0+\rows*\ch});
}

\foreach \i in {1,...,11} {
  \draw[thin, gray!60] ({\x0},{\y0+\i*\ch}) -- ({\x0+\cols*\cw},{\y0+\i*\ch});
}

%\draw[thick] (\x0,\y0) rectangle ({\x0+\cols*\cw},{\y0+\rows*\ch});
% \draw[very thick, gray!75] ({\x0+4*\cw},\y0) -- ({\x0+4*\cw},{\y0+\rows*\ch});
% \draw[very thick, gray!75] ({\x0+8*\cw},\y0) -- ({\x0+8*\cw},{\y0+\rows*\ch});
% \draw[very thick, gray!75] ({\x0+12*\cw},\y0) -- ({\x0+12*\cw},{\y0+\rows*\ch});

% labels / bars / LPRS with larger spacing to avoid overlap
\node[anchor=west] at (\x0,{\gridtop+0.86}) {PRS};
% \node[anchor=west] at ({\x0+4*\cw},{\gridtop+0.86}) {gNB 2};
% \node[anchor=west] at ({\x0+8*\cw},{\gridtop+0.86}) {gNB 3};
% \node[anchor=west] at ({\x0+12*\cw},{\gridtop+0.86}) {unused};
\node[anchor=west] at ({\x0+8*\cw},{\gridtop+0.86}) {unused};

\fill[usrpgreen!30]   (\x0, {\gridtop+0.58}) rectangle ({\x0+3*\cw}, {\gridtop+0.70});
% \fill[usrppurple!85] ({\x0+4*\cw}, {\gridtop+0.58}) rectangle ({\x0+8*\cw}, {\gridtop+0.70});
% \fill[usrpgreen!88] ({\x0+8*\cw}, {\gridtop+0.58}) rectangle ({\x0+12*\cw}, {\gridtop+0.70});
% \fill[gray!18]       ({\x0+12*\cw}, {\gridtop+0.58}) rectangle ({\x0+14*\cw}, {\gridtop+0.70});
\fill[gray!18]       ({\x0+3*\cw}, {\gridtop+0.58}) rectangle ({\x0+14*\cw}, {\gridtop+0.70});

\draw[decorate,decoration={brace,amplitude=2.5pt}] (\x0,{\gridtop+0.10}) -- ({\x0+3*\cw},{\gridtop+0.10});
\node[font=\small, anchor=west] at (\x0,{\gridtop+0.38}) {$L_{\mathrm{PRS}}=3$};

% \draw[decorate,decoration={brace,amplitude=2.5pt}] ({\x0+4*\cw},{\gridtop+0.10}) -- ({\x0+8*\cw},{\gridtop+0.10});
% \node[font=\small, anchor=west] at ({\x0+4*\cw},{\gridtop+0.38}) {$L_{\mathrm{PRS}}=4$};
% \draw[decorate,decoration={brace,amplitude=2.5pt}] ({\x0+8*\cw},{\gridtop+0.10}) -- ({\x0+12*\cw},{\gridtop+0.10});
% \node[font=\small, anchor=west] at ({\x0+8*\cw},{\gridtop+0.38}) {$L_{\mathrm{PRS}}=4$};
% \draw[decorate,decoration={brace,amplitude=2.5pt}] ({\x0+12*\cw},{\gridtop+0.10}) -- ({\x0+14*\cw},{\gridtop+0.10});

% axes / ticks
\node[rotate=90] at (0.06,{\y0+2.15}) {$k$};
\node at ({\x0+7*\cw},{\y0-0.58}) {$l$};
\foreach \c in {1,...,14} {
  \node[font=\scriptsize] at ({\x0+(\c-0.5)*\cw}, {\y0-0.28}) {\c};
}
\foreach \r/\lab in {1/1,6/6,12/12} {
  \node[font=\scriptsize] at ({0.32},{\y0+(\r-0.5)*\ch}) {\lab};
}

% neat legend, slightly higher and more even
% \matrix[matrix of nodes,
%         anchor=west,
%         row sep=0pt,
%         column sep=8pt,
%         nodes={anchor=west, inner sep=0pt, outer sep=0pt},
%        ] at (0.05,0.92) (leg) {
%    \node[draw=none, fill=usrpblue!85, minimum width=0.22cm, minimum height=0.16cm] {}; &
%    \node[anchor=west] {gNB 1}; &
%    \node[draw=none, fill=usrppurple!85, minimum width=0.22cm, minimum height=0.16cm] {}; &
%    \node[anchor=west] {gNB 2}; &
%    \node[draw=none, fill=usrpgreen!88, minimum width=0.22cm, minimum height=0.16cm] {}; &
%    \node[anchor=west] {gNB 3}; &
%    \node[draw=none, fill=gray!18, minimum width=0.22cm, minimum height=0.16cm] {}; &
%    \node[anchor=west] {unused}; \\
% };

% ============================================================
% Right: processing with stacked Y and H blocks
% ============================================================
\begin{scope}[shift={(6.60,1.00)}]

% removed node text "receiver processing"

% stacked Y blocks
\node[draw,  rectangle, minimum width=1.25cm, minimum height=0.62cm, fill=usrpblue!30, align=center, anchor=west] (y3) at (1.85,5.5)
  {$Y_{a=2}$};
% \node[draw,  rectangle, minimum width=1.95cm, minimum height=0.62cm, fill=usrppurple!30, align=center, anchor=west] (y2) at (0.65,6.66)
%   {$Y_{a=2}$};
\node[draw,  rectangle, minimum width=1.25cm, minimum height=0.62cm, fill=usrpblue!30, text=black, align=center, anchor=west] (y1) at (0.45,5.5)
  {$Y_{a=1}$};

% LS formula block
\node[draw, rounded corners, minimum width=3cm, minimum height=1.10cm, fill=yellow!18, align=center, anchor=west] (ls) at (0.15,3.80)
  {$\widehat{H}_{a}[k,l] = \dfrac{Y_{a}[k,l]}{X_{}[k,l]}$};

% define centered targets along top of LS block
\coordinate (lsTopL) at ($(ls.north west)!0.3!(ls.north east)$);
\coordinate (lsTopC) at ($(ls.north west)!0.50!(ls.north east)$);
\coordinate (lsTopR) at ($(ls.north west)!0.78!(ls.north east)$);

% define centered targets along bottom of LS block
\coordinate (lsBotL) at ($(ls.south west)!0.3!(ls.south east)$);
\coordinate (lsBotC) at ($(ls.south west)!0.53!(ls.south east)$);
\coordinate (lsBotR) at ($(ls.south west)!0.76!(ls.south east)$);

% stacked H outputs
\node[draw,  rectangle, minimum width=1.20cm, minimum height=0.66cm, fill=usrpblue!30, align=center, anchor=west] (h3) at (1.85,2.26)
  {$\widehat{H}_{a=2}$};
% \node[draw,  rectangle, minimum width=2.20cm, minimum height=0.66cm, fill=usrppurple!30, align=center, anchor=west] (h2) at (0.65,2.44)
%   {$\widehat{H}_{g=2}$};
\node[draw,  rectangle, minimum width=1.20cm, minimum height=0.66cm, fill=usrpblue!30, text=black, align=center, anchor=west] (h1) at (0.45,2.26)
  {$\widehat{H}_{a=1}$};

\begin{scope}%[on background layer]
\draw[->, thick] (lsBotL) -- (h1.north);
% \draw[->, thick] (lsBotC) -- (h2.north);
\draw[->, thick] (lsBotR) -- (h3.north);
\end{scope}

% arrows from center to center-ish points
\draw[->, thick] (y1.south) -- (lsTopL);
% \draw[->, thick] (y2.south) -- (lsTopC);
\draw[->, thick] (y3.south) -- (lsTopR);
\end{scope}
% \draw[->, thick] ({\x0+\cols*\cw},{\y0+\rows*\ch-0.8}) -- (y1.south west) ;
% \draw[->, thick] ({\x0+\cols*\cw},{\y0+\rows*\ch-0.8}) -- (y2.south west) ;
% \draw[->, thick] ({\x0+\cols*\cw},{\y0+\rows*\ch-1.2}) -- (y3.south west) ;
\end{tikzpicture}
}
\caption{Representation of \ac{5g} frame (20 slots at numerology 1), \ac{prs} comb-2 symbol allocation in a slot, and \ac{CSI} extraction. 
%Three \acp{gnb} while each occupies four \ac{ofdm} symbols within a slot with comb-2.
}
\label{fig:5g_prs}
\end{figure}
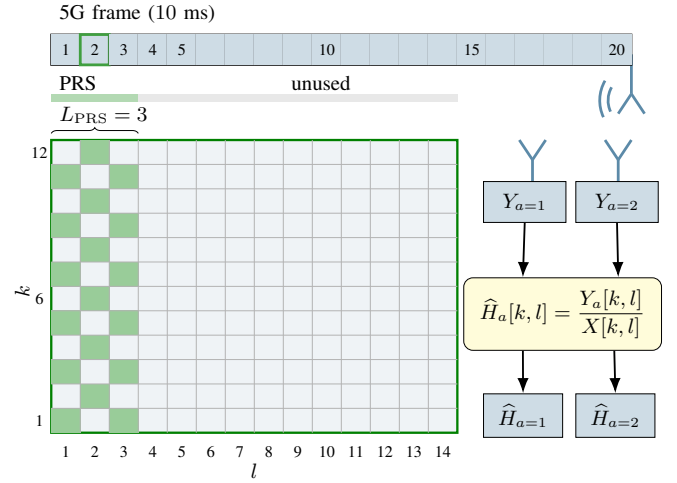

\subsection{CSI Estimation and Preprocessing}

% \textcolor{red}{why $Y$ and $W$ are not dependent on $g$?}
% \textcolor{orange}{I forgot to add a sentence saying we have a 3x vector of $Y$. And also $W$.}

%{For each demodulated \ac{ofdm} pilot symbol, the received resource-grid sample at receive antenna $a\in\{1,2\}$, \ac{ofdm}-symbol index $l$, and subcarrier $k=1,\ldots,N_{\mathrm{K}}$ is denoted by $Y_{a}[k,l]$. 
The received pilot sample at antenna $a$, symbol $l$ and subcarrier $k$ is
\begin{align}
Y_{a}[k,l]=
X^{\mathrm{PRS}}[k,l]H_{a}[k,l]+W_{a}[k,l],
\end{align}
where $X^{\mathrm{PRS}}[k,l]$ is the known transmitted \ac{prs} symbol, $H_{a}[k,l]$ is the downlink channel frequency response from the \ac{gnb} to receive antenna $a$, and $W_{a}[k,l]$ denotes noise and residual interference.

The symbol-level \ac{LS} estimate of the \ac{prs}-based \ac{CSI} is obtained independently for each receive antenna as
\begin{align}
\widehat{H}_{a}^{\mathrm{}}[k,l]=
\frac{Y_{a}[k,l]}{X^{\mathrm{PRS}}[k,l]}.
\end{align}
Since the \ac{prs} comb pattern only occupies a subset of subcarriers, the symbol-level estimates are first interpolated along the subcarrier dimension to obtain $\widehat{H}_{a}^{\mathrm{int}}[k,l]$, for $k=1,\ldots,N_{\mathrm{K}}$.

In our processing, the \ac{CSI} estimates are also organized at the slot level. 
For each slot $s$, the interpolated \ac{prs}-based symbol-level estimates $\widehat{H}_{a}^{\mathrm{int}}[k,l]$ are averaged over the $L_{\mathrm{PRS}}$ \ac{prs} \ac{ofdm} symbols to obtain the slot-level estimate $\widehat{H}_{a}[k,s]$.

\subsection{CSI Preprocessing Steps}
\label{subsubsec:csi_preprocessing}

The estimated \ac{CSI} comprises both amplitude and phase information that can be directly provided to the \ac{DL} model. In addition, we evaluate two commonly adopted pre-processing operations: temporal \ac{sg} filtering of the amplitude and \ac{lpr} \cite{diaz2023phase}.
% The amplitude component is primarily affected by path-loss, shadowing, reflections, and human-induced variations in the multipath structure.
% The phase, on the other hand, is generally more sensitive to small displacements and fine-grained movements, as even sub-wavelength changes in the paths can produce significant phase variations. 

% \sout{
% To reduce high-frequency fluctuations in the amplitude values, a \ac{sg} filter is applied along the frame dimension with polynomial order $2$ and a window length of $21$}
% \textcolor{red}{motivation for parameters selection?} \textcolor{orange}{I take every five \ac{5g} frames with $100$ slots then process the \ac{sg} filter while the half-window length $l$ is selected as $0.1$ times the input-vector length. The complete filter window therefore contains $2l+1$ samples \cite{diaz2023phase}}. 

As concerns the amplitude, % to reduce high-frequency fluctuations, a \ac{sg} filter is applied along the temporal dimension.
% according to \cite{diaz2023phase},
a polynomial order of two and a window length equal to approximately $0.2$ times the input-vector length were found to provide the best performance. Following the same empirical principle, we treat $400$ consecutive slot-level \ac{CSI} estimates as one sequence and use a second-order polynomial with a window length of $81$ samples.

The phase branch requires additional processing because the measured \ac{CSI} phase is not only determined by the physical propagation channel \cite{diaz2023phase}.
It can also contain phase errors caused by hardware and synchronization effects. 
Previous \ac{CSI} phase studies have shown that raw phase can contain random phase offsets and linear phase distortions, and that a \ac{lpr} can be used to calibrate the measured phase~\cite{PhaseFi}. 
In addition, a sampling timing offset introduces a phase term that is linear over the subcarrier domain~\cite{kotaruSpotFiDecimeterLevel2015}. 
Therefore, the measured phase is modeled as
\begin{align}
\Phi_{a}[k,	s]
=
\Phi^{\mathrm{ch}}_{a}[k,s]+\theta_{a}[s]+\gamma_{a}[s]k+\xi_{a}[k,s],
\label{eq:csi_phase_error_model}
\end{align}
where $\Phi^{\mathrm{ch}}_{a}[k,s]$ is the channel-related phase, $\theta_{a}[s]$ an unknown common phase offset, 
% \textcolor{red}{this should be the residual bias that is not corrected by CFO removal in the synch procedure, but we're not talking about it so maybe it is fine like this}, 
% \textcolor{orange}{as a common bias along the subcarrier axis?}\textcolor{red}{yes}
$\gamma_{a}[s]k$ the linear phase distortion across subcarriers, and $\xi_{a}[k,s]$ a residual random phase noise.

The aim of the preprocessing phase is to reduce the non-channel phase terms in~\eqref{eq:csi_phase_error_model}.
For each slot $s$, the unwrapped phase is fitted by the linear error component
$\widehat{\Phi}^{\mathrm{lin}}_{a}[k,s]
=
\widehat{\theta}_{a}[s]
+
\widehat{\gamma}_{a}[s]k,$
where the estimated offset and slope are obtained by
\begin{align}
\left(
\widehat{\theta}_{a}[s],
\widehat{\gamma}_{a}[s]
\right)
=
\arg\min_{\theta,\gamma}
\sum_{k=1}^{N_{\mathrm{K}}}
\left(
{\Phi}_{a}[k,s]-\theta-\gamma k
\right)^2 .
\label{eq:linear_phase_fit_coefficients}
\end{align}
The fitted linear component can be removed by subtraction:
\begin{align}
\widetilde{\Phi}_{a}[k,s]
=
{\Phi}_{a}[k,s]
-
\widehat{\Phi}^{\mathrm{lin}}_{a}[k,s].
\label{eq:linear_phase_removal}
\end{align}
% Therefore, the subcarrier-linear phase distortion is reduced, while the channel-related residual phase is preserved.
% In addition, after linear phase correction, a Savitzky--Golay filter can be applied along the temporal dimension to smooth residual phase fluctuations, as used in~\cite{diaz2023phase}. 
% After \ac{lpr}, the corrected phase becomes
When \ac{lpr} is enabled, the phase representation is obtained as
$\widetilde{\Phi}_{a}[k,s]\approx\Phi^{\mathrm{ch}}_{a}[k,s]+\xi_{a}[k,s]$, and the phase-corrected \ac{CSI} is reconstructed as
\begin{align}
\widetilde{H}_{a}[k,s]
=
\left|
\widehat{H}_{a}[k,s]
\right|
e^{j\widetilde{\Phi}_{a}[k,s]}.
\label{eq:phase_corrected_csi}
\end{align}

% \textcolor{red}{No sense here. here it is channel processing. the camera adaptation is only required for the evaluation. here the focus must stay on the wireless. The next paragraph is pure report style, no scientific.}
% \textcolor{orange}{moved to section 4}
% \sout{The camera provides pose labels at $8.66~\mathrm{Hz}$ with one camera frame every $115~${ms}. The keypoints are linearly interpolated with a factor of 4 between camera frames to achieve a frame of $28.9$\,{ms}. For numerology $\mu=1$, one \ac{5g} radio frame has duration $10~\mathrm{ms}$. After \ac{LS} estimation, several frames are jumped because some synchronization blocks are not recognized and downgrades the \ac{5g} radio samples with a factor of $0.7$ to duration about $14.3$\,ms. Therefore, one camera frame corresponds to about $N^t_\mathrm{slot} = 2 \times 20 = 40$ slots, with $N^t_\mathrm{slot}$ denoting number of slots per camera frame $t$. These slots are aligned with the camera frame, then taken into one \ac{CSI} sample $t$.  }

Using the pre-processed \ac{CSI}, the matrix is then constructed by concatenating along the subcarrier domain and slot domain with 
\begin{align} 
\widetilde{\mathbf{H}}^t_{a} &= 
\begin{bmatrix} 
\widetilde{H}_{a}[1,1] &  \cdots & \widetilde{H}_{a}[1,N^t_\mathrm{slot}] 
\\ 
\vdots  & \ddots & \vdots 
\\
\widetilde{H}_{a}[N_{\mathrm{K}},1] & \cdots  & \widetilde{H}_{a}[N_{\mathrm{K}},N^t_\mathrm{slot}] 
\end{bmatrix} 
\in \mathbb{C}^{N_{\mathrm{K}}\times N^t_\mathrm{slot}}, 
\label{eq:csi_matrix_per_antenna} 
\end{align}
with $N^t_\mathrm{slot}$ denoting number of slots in one \ac{CSI} sample $t$.
After concatenating also the antenna domain $a$, the \ac{CSI} tensor used as the learning input has size
$\widetilde{\mathbf{H}}^t_{} \in \mathbb{C}^{2\times N_{\mathrm{K}}\times N^t_\mathrm{slot}}$.
% where the three dimensions correspond to the three \acp{gnb}, the subcarrier dimension, and the aligned slot-level time dimension, respectively.
The amplitude and the unwrapped phase along the subcarriers are obtained from the \ac{CSI} tensor as
$\left[\left|\widetilde{\mathbf{H}}^t_{} \right|\, \angle\widetilde{\mathbf{H}}^t_{}\right]$
These two tensors are then used as selectable input branches for the \ac{DL} model estimating the pose.

\section{Pose Prediction Models and Pipeline}
\label{sec:baseline}

% Preamble requirements:
% \usepackage{tikz}
% \usepackage{graphicx}
% \usetikzlibrary{positioning, shapes.geometric, arrows.meta, calc, decorations.pathreplacing}

\begin{figure}[!tb]
\centering
\begin{adjustbox}{trim={16pt 1pt 26pt 1pt},clip}

\resizebox{1.16\linewidth}{!}{%
% defineblock: no-op environment for code navigation.
% Search for: \begin{defineblock}{...}
\newenvironment{defineblock}[1]{%
  \def\currentDefineBlock{#1}%
  \typeout{TikZ defineblock: #1}%
}{}

\begin{tikzpicture}[
  x=1cm,y=-1cm,
  font=\sffamily\normalsize,
  >=Stealth,
  line cap=round,
  line join=round,
  % csiFront/.style={draw=usrpblue,fill=usrpblue!30,very thick,minimum width=1.1cm,minimum height=0.9cm,align=center,font=\sffamily\large},
csiFront/.style={draw=usrpblue,fill=usrpblue!30,very thick,minimum width=1.6cm,minimum height=1.0cm,align=center,font=\sffamily\Large},
  csiBackP/.style={draw=usrppurple,fill=usrppurple!30,thick,minimum width=1.6cm,minimum height=1.0cm},
  csiBackO/.style={draw=usrpgreen,fill=usrpgreen!30,thick,minimum width=1.1cm,minimum height=0.9cm},
  usrp/.style={draw=#1,rounded corners=4pt,very thick,fill=#1!8,minimum width=1.4cm,minimum height=0.8cm,text width=1.3cm,align=center,font=\sffamily\large,text=#1},
  net/.style={draw=largeblockedge,fill=largeblockedge!5,very thick,minimum height=0.6cm,align=center,font=\sffamily\normalsize},
  ampblk/.style={draw=ampedge,fill=ampfill,very thick,minimum width=1.8cm,minimum height=0.6cm,align=center,font=\sffamily\normalsize},
  arrow/.style={-Stealth,very thick,draw=black},
  thinArrow/.style={-Stealth,thick,draw=black},
  dim/.style={<->,thick,draw=black!80}
]

%\definecolor{usrpblue}{RGB}{64,113,175}
%\definecolor{usrppurple}{RGB}{124,95,158}
%\definecolor{usrporange}{RGB}{239,137,68}
\definecolor{usrpblue}{rgb}{0.36, 0.54, 0.66}
\definecolor{usrppurple}{rgb}{0.36, 0.54, 0.66}
% \definecolor{usrppurple}{rgb}{0.9, 0.17, 0.31}
\definecolor{usrpgreen}{rgb}{0.0, 0.5, 0.0}

%\definecolor{dtfill}{RGB}{225,243,216}
%\definecolor{dtedge}{RGB}{21,137,72}
\definecolor{dtfill}{rgb}{0.7, 0.75, 0.71}
\definecolor{dtedge}{rgb}{0.43, 0.5, 0.5}
\definecolor{ampfill}{RGB}{242,242,232}
\definecolor{ampedge}{RGB}{40,40,40}

\definecolor{largeblockedge}{rgb}{0,0,0}
\definecolor{poseorange}{RGB}{229,92,10}
\definecolor{posered}{rgb}{1.0, 0.0, 0.0}
\pgfdeclarelayer{bg}
\pgfsetlayers{bg,main}

% overall picture bounding box + requested corner coordinates
\coordinate (figNW) at (-0.8,-0.2);
\coordinate (figNE) at (15.6,-0.2);
\coordinate (figSW) at (-0.8,7.9);
\coordinate (figSE) at (15.6,7.9);
\path[use as bounding box] (figNW) rectangle (figSE);

% --------------------- helper macros ---------------------
\newcommand{\tripstack}[4]{%
  \begin{scope}[shift={(#1,#2)}]
    % Equal-size layers with equal diagonal offsets.
    % \node[csiBackO] (#3o) at (0.4,-0.4) {};
    % \node[csiBackP] (#3p) at (0.2,-0.2) {};
    \node[csiFront] (#3b) at (0.0,0.0) {#4};
  \end{scope}%
}

\newcommand{\tripstackleft}[4]{%
  \begin{scope}[shift={(#1,#2)}]
    % Left stack: back layers go toward upper-left instead of upper-right.
    % \node[csiBackO] (#3o) at (-0.4,-0.4) {};
    \node[csiBackP] (#3p) at (-0.2,-0.2) {};
    \node[csiFront] (#3b) at (0.0,0.0) {#4};
  \end{scope}%
}

\newcommand{\hstack}[4]{%
  \begin{scope}[shift={(#1,#2)}]
    % Equal-size H_tild layers with the same diagonal offsets as CSI.
    % \node[csiBackO,minimum width=1.6cm,minimum height=1.0cm] (#3o) at (-0.4,-0.4) {};
    \node[csiBackP,minimum width=1.6cm,minimum height=1.0cm] (#3p) at (-0.2,-0.2) {};
    \node[csiFront,minimum width=1.6cm,minimum height=1.0cm,font=\sffamily\Large] (#3b) at (0.0,0.0) {#4};
  \end{scope}%
}

\newcommand{\txantennaLeft}[3]{%
  \begin{scope}[shift={(#1,#2)},color=#3]
    \draw[very thick] (0,0) -- (0,-0.6);
    \draw[very thick] (0,-0.6) -- (-0.2,-0.9);
    \draw[very thick] (0,-0.6) -- (0.2,-0.9);
    \draw[very thick] (-0.3,-0.5) arc[start angle=210,end angle=155,radius=0.4];
    \draw[very thick] (-0.4,-0.6) arc[start angle=210,end angle=155,radius=0.6];
  \end{scope}%
}

\newcommand{\txantennaRight}[3]{%
  \begin{scope}[shift={(#1,#2)},color=#3]
    \draw[very thick] (0,0) -- (0,-0.6);
    \draw[very thick] (0,-0.6) -- (-0.2,-0.9);
    \draw[very thick] (0,-0.6) -- (0.2,-0.9);
    \draw[very thick] (0.3,-0.5) arc[start angle=-30,end angle=25,radius=0.4];
    \draw[very thick] (0.4,-0.6) arc[start angle=-30,end angle=25,radius=0.6];
  \end{scope}%
}

\newcommand{\rxantenna}[2]{%
  \begin{scope}[shift={(#1,#2)},color=usrpblue]
    \draw[very thick] (0,0) -- (0,-0.6);
    \draw[very thick] (0,-0.6) -- (-0.2,-0.9);
    \draw[very thick] (0,-0.6) -- (0.2,-0.9);
    \draw[very thick] (0.5,0) -- (0.5,-0.6);
    \draw[very thick] (0.5,-0.6) -- (0.3,-0.9);
    \draw[very thick] (0.5,-0.6) -- (0.7,-0.9);
  \end{scope}%
}

\newcommand{\usrptx}[4]{%
  \node[usrp=#1] (#2) at (#3,#4) {USRP\\Tx};
}

\newcommand{\drawcoco}[4]{%
\begin{scope}[shift={(#1,#2)},scale=#3,color=#4,line cap=round,line join=round]
  \coordinate (nose) at (0,0);
  \coordinate (leye) at (-0.2,-0.1);
  \coordinate (reye) at (0.2,-0.1);
  \coordinate (lear) at (-0.3,0.0);
  \coordinate (rear) at (0.3,0.0);
  \coordinate (lsho) at (-0.8,0.7);
  \coordinate (rsho) at (0.8,0.7);
  \coordinate (lelb) at (-1.2,1.5);
  \coordinate (relb) at (1.2,1.5);
  \coordinate (lwri) at (-1.3,2.2);
  \coordinate (rwri) at (1.3,2.2);
  \coordinate (lhip) at (-0.5,2.1);
  \coordinate (rhip) at (0.5,2.1);
  \coordinate (lkne) at (-0.7,3.3);
  \coordinate (rkne) at (0.7,3.3);
  \coordinate (lank) at (-0.7,4.2);
  \coordinate (rank) at (0.7,4.2);
  \draw[thick] (lear)--(leye)--(nose)--(reye)--(rear);
  \draw[thick] (lear)--(lsho);
  \draw[thick] (rear)--(rsho);
  \draw[thick] (lsho)--(rsho);
  \draw[thick] (lsho)--(lelb)--(lwri);
  \draw[thick] (rsho)--(relb)--(rwri);
  \draw[thick] (lsho)--(lhip)--(rhip)--(rsho);
  \draw[thick] (lhip)--(lkne)--(lank);
  \draw[thick] (rhip)--(rkne)--(rank);
  \foreach \p in {nose,leye,reye,lear,rear,lsho,rsho,lelb,relb,lwri,rwri,lhip,rhip,lkne,rkne,lank,rank}
    \fill (\p) circle (0.9pt);
\end{scope}%
}

%%%%%%%%%%%%%%%%%%% --------------------- diagram blocks ---------------------
\begin{defineblock}{01-csi-input-stacks}
\tripstackleft{1.7}{0.9}{csiL}{$\widehat{H_1}$}
% \tripstack{2.9}{0.9}{csiR}{$\widehat{H_2}$}
\end{defineblock}

\node[anchor=south,text=black,font=\sffamily\large] at (3.5,0.8) {LS};

\begin{defineblock}{02-csi-ratio-and-h-tild}
\node[draw=black,fill=white,circle,very thick,minimum size=0.7cm,font=\sffamily\Large] (ratio) at (1.7,2.5) {$-$};
\node[anchor=south,text=black,font=\sffamily\large] at (4.1,3.0) {w/o Pre-Processing};

% CSI -> ratio: one-bend polylines.
% Each arrow starts from its colored CSI layer bottom, goes down, then enters the circle side.
% The final entering segments are horizontal/parallel.
\begin{pgfonlayer}{bg}
% left/right circle-wall targets, top-to-bottom
\coordinate (ratioL1) at (ratio.150);
\coordinate (ratioL2) at (ratio.180);
\coordinate (ratioL3) at (ratio.210);
\coordinate (ratioR1) at (ratio.30);
\coordinate (ratioR2) at (ratio.0);
\coordinate (ratioR3) at (ratio.330);

% bend points: same x as the start, same y as the target
% so departure is vertical and entry is horizontal
\coordinate (csiL1b) at ($(csiLb.south |- ratioL1)$);
% \coordinate (csiL2b) at ($(csiLp.south |- ratioL2)$);
% \coordinate (csiL3b) at ($(csiLo.south |- ratioL3)$);
% \coordinate (csiR1b) at ($(csiRb.south |- ratioR1)$);
% \coordinate (csiR2b) at ($(csiRp.south |- ratioR2)$);
% \coordinate (csiR3b) at ($(csiRo.south |- ratioR3)$);

% \draw[-Stealth,thick,draw=usrpblue]   (csiLb.south) -- (csiL1b) -- (ratioL1);
\draw[-Stealth,thick,draw=usrpblue]   (csiLb.south) -- (ratio.90);
% \draw[-Stealth,thick,draw=usrppurple] (csiLp.south) -- (csiL2b) -- (ratioL2);
% \draw[-Stealth,thick,draw=usrpgreen] (csiLo.south) -- (csiL3b) -- (ratioL3);
% \draw[-Stealth,thick,draw=usrpblue]   (csiRb.south) -- (csiR1b) -- (ratioR1);
% \draw[-Stealth,thick,draw=usrppurple] (csiRp.south) -- (csiR2b) -- (ratioR2);
% \draw[-Stealth,thick,draw=usrpgreen] (csiRo.south) -- (csiR3b) -- (ratioR3);
\end{pgfonlayer}

\hstack{1.9}{4.2}{hF}{$\widetilde{H}$}

% aligned arrows from ratio to H_tild
\coordinate (ratioOutB) at (ratio.295);
\coordinate (ratioOutP) at (ratio.270);
\coordinate (ratioOutO) at (ratio.245);

\coordinate (hF) at ($(hFb.north west)!0.4!(hFb.north east)$);
\draw[-Stealth,very thick,draw=usrpblue]   (ratioOutP) -- (hF);
% \draw[-Stealth,very thick,draw=usrppurple] (ratioOutP) -- (hFp.north);
% \draw[-Stealth,very thick,draw=usrpgreen] (ratioOutO) -- (hFo.north);

% dimension arrows + labels bound to H_tild
% \draw[dim] (0.5,3.4) -- (0.5,4.4);
% \node[anchor=east,text=black!85,font=\sffamily\large] at (0.3,3.9) {$N_{\mathrm{Rx}}=3$};

\draw[dim] (3.0,3.7) -- (3.0,4.7);
\node[anchor=west,text=black!85,font=\sffamily\Large] at (3.0,4.0) {$N_\mathrm{K} = 612$};

\draw[dim] (1.1,4.9) -- (2.7,4.9);
\node[anchor=west,text=black!85,font=\sffamily\Large] at (2.6,5.1) {$N^t_\mathrm{slot} = 40$};

\begin{pgfonlayer}{bg}
\draw[-Stealth,very thick,draw=usrpblue]   (hFb.south) -- (1.9,5.6);
% \draw[-Stealth,very thick,draw=usrppurple] (hFp.south) -- (1.7,5.6);
% \draw[-Stealth,very thick,draw=usrpgreen] (hFo.south) -- (1.5,5.6);
\end{pgfonlayer}
\end{defineblock}

\begin{defineblock}{03-usrp-rx}
\node[draw=usrpblue,rounded corners=4pt,very thick,fill=white,minimum width=1.4cm,minimum height=0.8cm,text width=1.3cm,align=center,font=\sffamily\large,text=usrpblue] (usrpRx) at (5.2,0.9) {USRP\\Rx};
\rxantenna{6.4}{0.9}
\draw[usrpblue,very thick] (usrpRx.east) -- (6.4,0.9) -- (6.9,0.9);
\draw[-Stealth,usrpblue,very thick] ($(usrpRx.west)+(0,0)$) -- ++(-1.8,0);
% \draw[-Stealth,usrpblue,very thick] ($(usrpRx.west)+(0,-0.2)$) -- ++(-1,0);
\end{defineblock}

% \begin{defineblock}{04-usrp-tx-c}
% \usrptx{usrpgreen}{usrpC}{5.2}{3.0}
% \draw[usrpgreen,very thick] (usrpC.east) -- (6.4,3.0);
% \txantennaRight{6.4}{3.0}{usrpgreen}
% \end{defineblock}

\begin{defineblock}{05-original-image}
\node[inner sep=0] (orig) at (9.3,2.0) {\includegraphics[width=3.5cm,trim={160 0 160 80},clip]{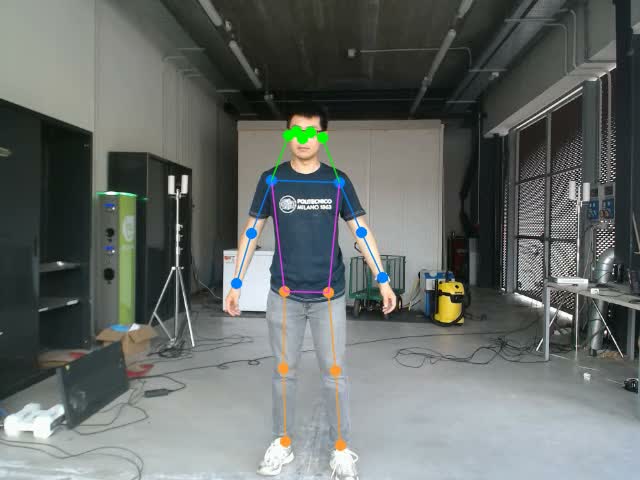}};
\end{defineblock}

% \begin{defineblock}{06-usrp-tx-a}
% \usrptx{usrpblue}{usrpA}{13.3}{0.9}
% \draw[usrpblue,very thick] (usrpA.west) -- (11.9,0.9);
% \txantennaLeft{11.9}{0.9}{usrpblue}
% \end{defineblock}

\begin{defineblock}{07-usrp-tx-b}
\usrptx{usrpblue}{usrpB}{13.3}{0.9}
\draw[usrpblue,very thick] (usrpB.west) -- (11.9,0.9);
\txantennaLeft{11.9}{0.9}{usrpblue}
\end{defineblock}

\begin{defineblock}{08-camera-yolo}
\coordinate (cam) at (9.2,4.3);
\draw[fill=black!85,draw=black!85] ($(cam)+(-0.2,0)$) -- ($(cam)+(0.2,0)$) -- ($(cam)+(0,0.3)$) -- cycle;
\draw[fill=black!85,draw=black!85] (9.0,4.6) rectangle (9.4,5.0);
\node[net,minimum width=1.2cm,minimum height=0.6cm,text=black] (yolo) at (10.9,4.8) {YOLO};
\draw[thinArrow] (9.4,4.8) -- (yolo.west);
\end{defineblock}

\begin{defineblock}{09-ground-truth-and-predicted-pose}
\draw[black!75,dashed,very thick,fill=white] (12.1,4.2) rectangle (14.1,5.7);
\node[anchor=north west,align=left,text=black,font=\sffamily\normalsize] at (12.0,4.2) {Ground\\Truth};
\drawcoco{13.6}{4.4}{0.3}{black!75}
\coordinate (gtYoloTarget) at (12.1,4.8);
\draw[thinArrow] (yolo.east) -- (gtYoloTarget);

\draw[black!75,dashed,very thick,fill=white] (12.1,5.8) rectangle (14.1,7.4);
\node[anchor=north west,align=left,text=black,font=\sffamily\normalsize] at (12.0,5.8) {Pred.\\Pose};
\drawcoco{13.6}{6.0}{0.3}{posered}
\coordinate (predMidLeft) at (12.1,6.6);
\end{defineblock}

\begin{defineblock}{10-dt-pose-pipeline}
    \fill[dtfill!20] (0.1,5.6) rectangle (11.8,7.5);
    \draw[dtedge,very thick] (0.1,5.6) rectangle (11.8,7.5);
    \node[anchor=north east,text=dtedge,font=\sffamily\large] at (11.7,5.7) {DT-Pose};
    
    % equalized semiblock boundaries, with 0.6 cm gap between adjacent parts
    \node[ampblk] (amp) at (1.1,6.2) {Amplitude};
    \node[ampblk] (pha) at (1.1,6.9) {Phase};
    \node[net,minimum width=0.9cm] (maeA) at (3.0,6.2) {MAE};
    \node[net,minimum width=0.9cm] (maeP) at (3.0,6.9) {MAE};
    \draw[arrow] (amp.east) -- (maeA.west);
    \draw[arrow] (pha.east) -- (maeP.west);
    
    \draw[largeblockedge,dashed,very thick] (4.0,5.8) rectangle (6.0,7.3);
    \node[anchor=north east,text=largeblockedge,font=\sffamily\small] at (6.0,5.8) {Fusion};
    \node[draw=largeblockedge,fill=largeblockedge!5,circle,very thick,minimum size=0.3cm] (fA) at (4.6,6.2) {};
    \node[draw=largeblockedge,fill=largeblockedge!5,circle,very thick,minimum size=0.3cm] (fP) at (4.6,6.9) {};
    \node[draw=largeblockedge,fill=largeblockedge!5,circle,very thick,minimum size=0.4cm] (fO) at (5.4,6.6) {};
    \draw[arrow,draw=black!70] (maeA.east) -- (fA.west);
    \draw[arrow,draw=black!70] (maeP.east) -- (fP.west);
    \draw[arrow,draw=black!70] (fA.east) -- (fO.150);
    \draw[arrow,draw=black!70] (fP.east) -- (fO.210);
    
    \node[net,minimum width=1.2cm,minimum height=0.7cm] (task) at (6.9,6.6) {Task-\\Prompt};
    \node[net,minimum width=1.0cm,minimum height=0.7cm] (gcn) at (8.5,6.6) {GCN};
    \node[net,minimum width=1.8cm,minimum height=0.7cm] (trans) at (10.6,6.6) {Transformer};
    \draw[arrow] (fO.east) -- (task.west);
    \draw[arrow] (task.east) -- (gcn.west);
    \draw[arrow] (gcn.east) -- (trans.west);
    \draw[thinArrow] (trans.east) -- (predMidLeft);
\end{defineblock}

\draw[yellow!90!orange, line width=2.2pt, rounded corners]
    (6.7,0.5) ellipse [x radius=0.7, y radius=0.65];

\draw[yellow!90!orange, line width=2.2pt]
    (7.8,0.9) ellipse [x radius=0.3, y radius=0.3];

% \draw[pink!90!orange, line width=2.2pt, rounded corners]
    % (11.9,0.4) ellipse [x radius=0.7, y radius=0.7];

% \draw[pink!90!orange, line width=2.2pt]
%     (10.9,2.9) ellipse [x radius=0.3, y radius=0.3];
\end{tikzpicture}
}
    % figure content

\end{adjustbox}

\caption{
Overview of the proposed DT-Pose framework for \ac{5g} \ac{CSI} based 2D human pose estimation using two \acp{USRP} with camera-based GT supervision. The 2-antenna receiver is highlighted on the left, the single antenna transmitter is behind the camera.
}
\label{fig:csi-dtpose}
\end{figure}
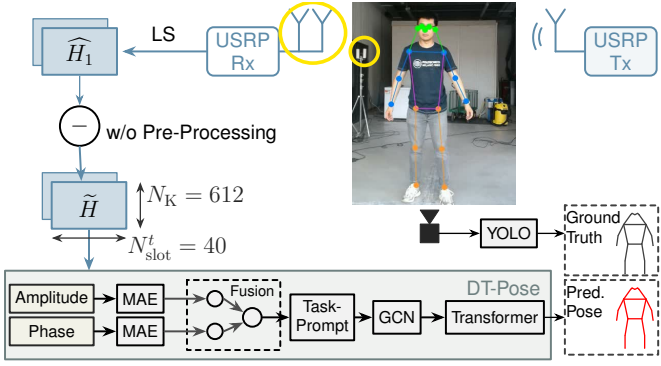

The full learning pipeline from \ac{prs}-based \ac{CSI} extraction to 2D pose prediction is illustrated in Fig.~\ref{fig:csi-dtpose}, while the experimental acquisition and \ac{CSI}-to-camera alignment are detailed in Section~\ref{sec:metrics}.
On the radio side, the two-antenna \ac{ue} estimates the \ac{CSI} from the received \ac{prs}, followed by preprocessing to obtain the amplitude and phase inputs, while the camera provides the corresponding 2D skeleton as ground truth.
The amplitude $|\widetilde{\mathbf{H}}^t|$ and phase $\angle\widetilde{\mathbf{H}}^t$ branches are independently encoded by the \ac{MAE} encoders, then fused and passed through the task-prompting block, \ac{GCN}, and Transformer modules to predict the final human pose.
% From the \ac{CSI} tensor, we use the amplitude branch $|\widetilde{\mathbf{H}}^t_{}|$ and the phase branch $\angle\widetilde{\mathbf{H}}^t_{}$.
%\textcolor{red}{the comments to the figure is too short} \textcolor{orange}{Corrected}

\subsection{DL Model}

DT-Pose \cite{chenRobustRealisticHuman2025} is adapted as a 2D pose-estimation backbone while keeping its main design unchanged. Following the original two-stage idea, we first pretrain the \ac{MAE}-\ac{ViT} encoder with a patch size of $(N_\mathrm{K}/N_\mathrm{PRB},4)$ in a self-supervised manner. Since the proposed input has two possible branches, the amplitude and phase branches are pretrained separately.
After pretraining, the \ac{MAE} encoders are used to extract branch-specific representations. For joint amplitude-phase training, the two branches are then combined by a learnable fusion:
\begin{align}
\mathbf{M}^{t}
&=
\alpha_1\widetilde{\mathbf{Z}}^t_\mathrm{amp} + \alpha_2\widetilde{\mathbf{Z}}^t_\mathrm{phase},
% &\sum_{i\in\{1,2\}}\alpha_i=1 ,
\label{eq:learnable_amp_phase_fusion}
\end{align}
with $\alpha_1+\alpha_2=1$, $\widetilde{\mathbf{Z}}^t_\mathrm{amp}$ denoting the amplitude branch after \ac{MAE} processing and $\widetilde{\mathbf{Z}}^t_\mathrm{phase}$ the phase after \ac{MAE}. 
%\textcolor{blue}{Here, $\alpha_1$ and $\alpha_2$ are initialized to 0.5.} \textcolor{red}{which are the final values?}
%\textcolor{orange}{Normally the amplitude related alpha becomes 0.51 and phase related alpha becomes 0.49. Also I realized that the learning rate is too small to change these values, although they mostly converge.}
%\textcolor{red}{so it is equal weight basically...}
%\textcolor{orange}{yes, so it means phase is relatively important. But the block of learnable fusion is a bad block for machine learning. It needs to be changed for the journal paper.}
For amplitude-only training, $\mathbf{M}^{t}=\widetilde{\mathbf{Z}}^t_\mathrm{amp}$. Therefore, the model input keeps the common form $\mathbf{M}^{t}\in\mathbb{R}^{2\times N_{\mathrm{K}}\times N_{\mathrm{s}}}$. 

The downstream DT-Pose decoder follows the original topology-aware design. It uses task prompts, \ac{GCN}, and transformer layers to model the structure of the human skeleton. The task prompt provides a learnable pose-aware query for the decoder. The \ac{GCN} layers then inject the human skeleton topology into the decoding process. The transformer layers capture long-range dependencies among human joints \cite{chenRobustRealisticHuman2025}.

The learning target is a 2D human skeleton, described according to the \ac{coco17} format. 
Accordingly, the graph-convolution part exploits the \ac{coco17} bone topology, adjacent joints and explicitly modeling the spatial relationship between connected body limbs considered during pose decoding. The final regression layer is modified to output 2D keypoints, where each output corresponds to the 2D coordinates of a specific anatomical landmark of the human body, such as the nose, shoulders, elbows, wrists, hips, knees, and ankles.

% MetaFi++ is also adapted to the same 2D output contract. The model receives the same input map $\mathbf{M}^{t}\in\mathbb{R}^{3\times N_{\mathrm{K}}\times N_{\mathrm{s}}}$. The three input channels correspond to the three \acp{gnb}. For comparison, we evaluate MetaFi++ with amplitude-only input and joint amplitude-phase input. In the joint setting, the same learnable amplitude-phase fusion is used before the fused map is passed to the MetaFi++ backbone.

\subsection{Loss Function}

For each camera frame, the model predicts 2D human skeleton keypoints
$\widehat{\mathbf{p}} \in \mathbb{R}^{J\times 2}$, where $J=17$ is the number of keypoints. The camera image resolution is $640 \times 480$. After extracting the 2D skeleton pixels, they are scaled by a factor to represent a meter scale, assuming the human is about $170$\,cm and is moving on a plane at a fixed camera depth. The corresponding \ac{GT} pose is denoted by $\mathbf{p}\in\mathbb{R}^{J\times 2}$.
The coordinate loss is the mean squared error of the keypoint estimate averaged over all frames:
\begin{align}
\mathcal{L}_{\mathrm{MSE}}
&=
\frac{1}{N_{\mathrm{t}}J}
\sum_{t=1}^{N_{\mathrm{t}}}
\sum_{j=1}^{J}
\left\|\widehat{\mathbf{p}}_{t,j} - \mathbf{p}_{t,j}\right\|_{2}^{2}.
\label{eq:pose_training_loss}
\end{align}
Here, the camera frame number $N_{\mathrm{t}}$ also denotes the number of training frames. All models are trained using the AdamW 
%\textcolor{purple}{(maybe this should also be an acronym)} \textcolor{orange}{All ML papers do not use acronym for Adam or AdamW, so I think it is fine e.g. \cite{chenRobustRealisticHuman2025}.}
optimizer with a learning rate of $10^{-4}$ and weight decay rate of $10^{-2}$. The batch size is chosen as $64$ and the model is trained for $200$ epochs.  
% The dataset is divided into training, validation, and test sets with a ratio of $7:2:1$. 
The dataset is divided into training, validation, and test subsets using non-overlapping temporal blocks with an approximate ratio of $7:2:1$ so that adjacent samples from same activity segments are not shared across subsets.
The final model is selected based on the lowest validation loss.

\begin{table}[t]
\centering
\caption{System parameters of the \ac{5g} \ac{prs}-based \ac{CSI} dataset
%for 2D pose estimation.
}
\label{tab:sys_params_prs_csi}
\begin{tabular}{lcc}
\toprule
System parameter & Symbol / key & Value \\
\midrule
%Carrier frequency & $f_c$ & $3.95~\mathrm{GHz}$ \\
% Number of \acp{gnb} & $N_{\mathrm{gNB}}$ & $4$ \\
% Number of \acp{ue} & $N_{\mathrm{UE}}$ & $1$ \\
Numerology & $\mu$ & $1$ \\
Bandwidth & $B$ & $20~\mathrm{MHz}$ \\
Number of \acp{prb} & $N_{\mathrm{PRB}}$ & $51$ \\
Number of subcarriers & $N_\mathrm{K}$ & $612$ \\
Subcarrier spacing & $\Delta f$ & $30~\mathrm{kHz}$ \\
\ac{prs} comb & $K_{\mathrm{COMB}}^{\mathrm{PRS}}$ & $2$ \\
\ac{prs} symbols & $L_{\mathrm{PRS}}$ & $3$ \\
Number Rx antennas & $N_{\mathrm{Rx}}$ & $2$ \\
Camera frame rate & $f_{\mathrm{cam}}$ & $8.66~\mathrm{Hz}$ \\
\ac{5g} frames per sample & $N_{\mathrm{f}}$ & $2$ \\
\ac{5g} slots per sample & $N^t_{\mathrm{slot}}$ & $40$ \\
Number of keypoints & $J$ & $17$ \\
Sensing area & -- & $5.3\times4.3~\mathrm{m}^{2}$ \\
\bottomrule
\end{tabular}
\end{table}

\subsection{Validation Metrics}
\label{sec:metrics}

The primary evaluation metrics are validation loss and \ac{PCK}.
The \ac{PCK} measures the average keypoint localization accuracy over all frames and all joints. It is defined as
\begin{align}
\mathrm{PCK}@\tau
&=
\frac{1}{N_{\mathrm{test}}J}
\sum_{t=1}^{N_{\mathrm{test}}}
\sum_{j=1}^{J}
\mathbf{1}\left(
\frac{
\left\|\widehat{\mathbf{p}}_{t,j} - \mathbf{p}_{t,j} \right\|_2
}{
\left\|\mathbf{p}_{t,\mathrm{ls}}-\mathbf{p}_{t,\mathrm{rh}}\right\|_2
}
\leq \tau
\right),
\label{eq:pck}
\end{align}
where $N_{\mathrm{test}}$ is the number of test frames, $J$ is the number of keypoints, and $\widehat{\mathbf{p}}_{t,j}$ and $\mathbf{p}_{t,j}$ are the predicted and \ac{GT} 2D coordinates of keypoint $j$ in frame $t$, respectively. The function $\mathbf{1}(\cdot)$ is a binary indicator. The normalization term $\left\|\mathbf{p}_{t,\mathrm{ls}}-\mathbf{p}_{t,\mathrm{rh}}\right\|_2$ is the distance between the left shoulder and the right hip \cite{zhou2023metafi++,chenRobustRealisticHuman2025}. In this work, we compute the \ac{PCK} at $\tau\in\{10\,\%,30\,\%,50\,\%\}$, meaning the percentage of keypoints with errors below the corresponding percentage of the reference torso distance.

\section{Model Evaluation}
\label{sec:results}

\subsection{Experimental Data Collection}

\begin{figure}[t]
\centering
\subfloat[Amplitude (normalized)]{
    \vspace{-8pt}
    \includegraphics[width=0.98\linewidth]{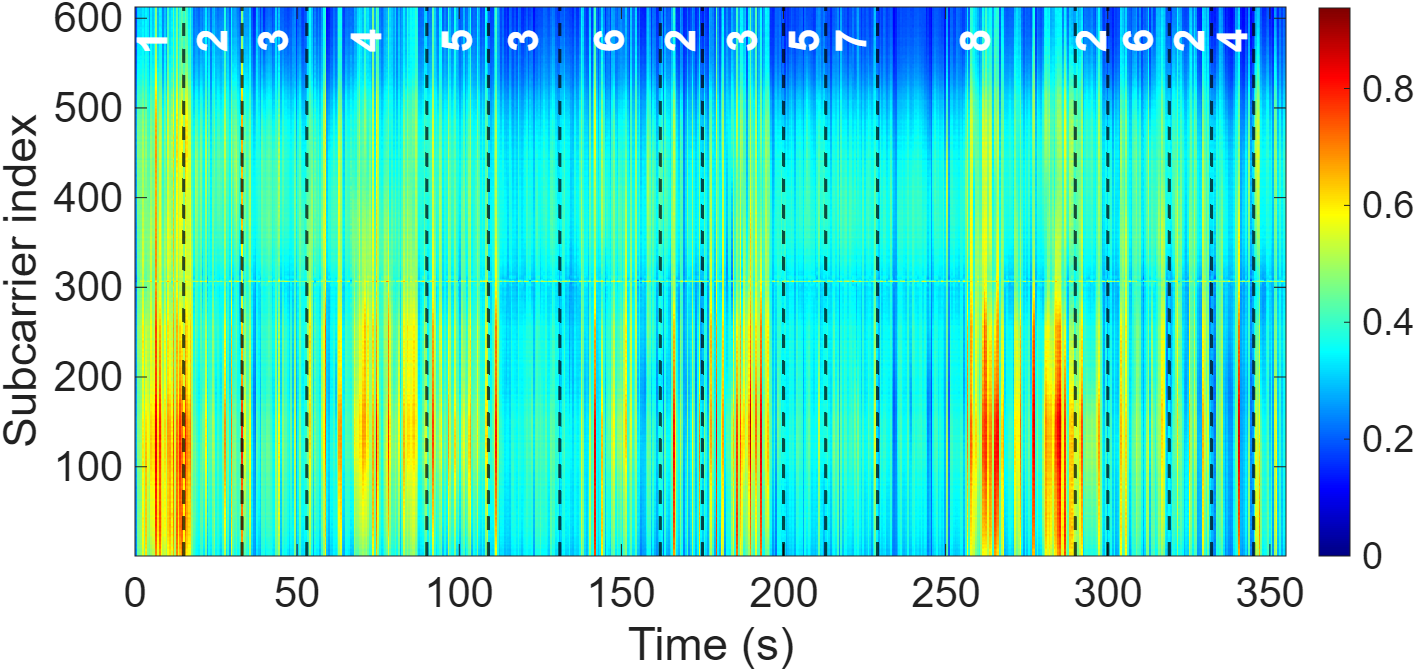}
    \label{fig:amp}
}
\\
\vspace{-7pt}
\subfloat[\text{Phase [degree]} ]{
    \includegraphics[width=0.99\linewidth]{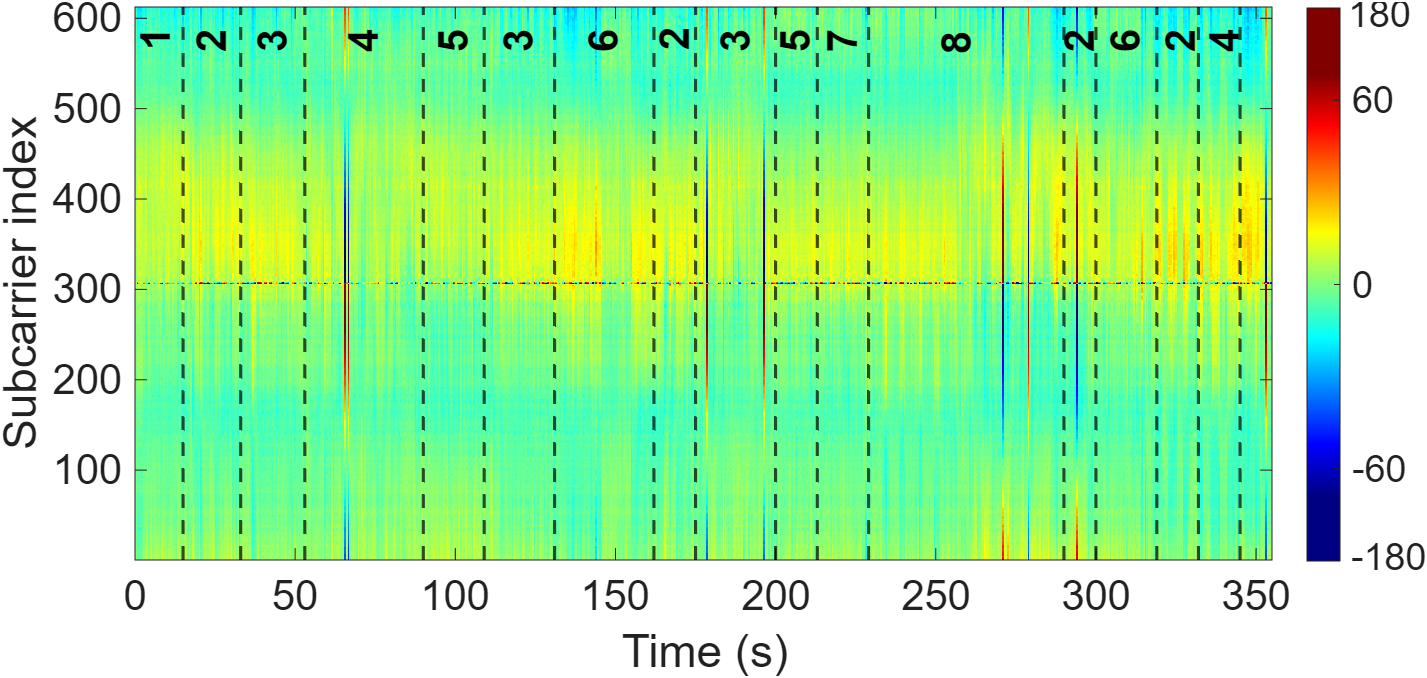}
    \label{fig:phase}
}
\caption{
% \textcolor{red}{try using jet colorbar. Why the colorbar ranges from -2 to 2? I'd use degree values, from -180 to 180.}
Preprocessed \ac{CSI} over the full activity sequence: 
(a) amplitude and (b) phase after linear component removal. 
The dashed vertical lines indicate the activity segments. The numbers indicate from 1 to 8: moving, shaking, bending, stretching, rotating, crouching, jumping, falling. 
}
\vspace{-7pt}
\label{fig:amp_phase_activity}
\end{figure}

% \textcolor{red}{never described the experiment, so it is impossible to understand what is the activity sequence. where do you describe the experiment? }
% \textcolor{orange}{done}
The measurements were collected in an indoor sensing area of $5.3\times4.3\,\mathrm{m}^{2}$, with
the parameters for the \ac{5g} and camera systems as reported in Table~\ref{tab:sys_params_prs_csi} and the \acp{USRP} share the same 10\,MHz reference synchronization signal
%\sout{with an Ettus Octoclock}
provided by an Ettus OctoClock-G CDA 2290. 
As illustrated in Fig.~\ref{fig:csi-dtpose}, one \ac{USRP} acts as \ac{gnb} with one transmit antenna and another \ac{USRP} acts as \ac{ue} equipped with two receive antennas. 
The yellow circle indicates the placement of the receive antennas. The transmit antenna was placed on the opposite edge of the sensing area, such that the human was located between the transmitter and receiver.
A single participant started approximately at the midpoint between the two devices and performed a continuous sequence of eight everyday activities. A set of $8$ different activities is considered, numbered as indicated in Fig. \ref{fig:amp_phase_activity} and in Table \ref{tab:activity_results_20mhz_joint}, with each activity repeated at least three times in each time interval. The human moved to the left side and returned to the initial position, then did repetitions of shaking hands, bending the body, stretching legs, rotation, crouching, jumping and falling. 

% \begin{enumerate}
%     \item moving to the left and returning to the initial position
%     \item shaking both hands
%     \item bending the upper body to the left, right, or front
%     \item stretching the legs
%     \item rotating the body
%     \item repetition of crouching
%     \item repetition of jumping
%     \item repetition of falling to the ground, standing up.
% \end{enumerate}

The camera provides pose labels at $f_\mathrm{cam} = 8.66~\mathrm{Hz}$ with one camera frame every $115~${ms}. The keypoints are linearly interpolated by a factor of 4 between camera frames to obtain a frame of $28.9$\,{ms}. For numerology $\mu=1$, one \ac{5g} radio frame has duration $10~\mathrm{ms}$. During the \ac{LS} estimation, several frames are lost because some synchronization blocks are not recognized, reducing the \ac{5g} radio samples by a factor of $0.7$. %to duration about $14.3$\,ms. 
Therefore, one camera frame corresponds to about $N^t_\mathrm{slot} = 2 \times 20 = 40$ slots. These slots are aligned with the camera frame, then taken into one \ac{CSI} sample $t$.   
With this setup, two experiments are recorded, denoted as $\beta_1$ or $\beta_2$. 
%\textcolor{blue}{
Experiment $\beta_1$ represents the primary measurement campaign with continuous activities, while experiment $\beta_2$ is recorded with faster activities and the recording is not continuous.
%}

% \textcolor{red}{It is not written which is the difference between the two tests}
% \textcolor{orange}{$\beta_2$ was recorded with a higher activity speed and contain less samples.}
% \textcolor{red}{well, given this comment, it is obvious for a reader to ask "what is $\beta_1$ then?}
% \textcolor{orange}{$\beta_1$ is the original test set for testing the feasibility of the pose est. pipeline......}

%in Table \ref{tab:input_bandwidth_results}. 
% The experiment $\beta_2$ with $40\,\mathrm{MHz}$ model is trained with only $4072$ samples. The sample number of $\beta_2$ is low and should therefore be interpreted as a preliminary evaluation. However, we compare this experiment with full $40\,\mathrm{MHz}$ and the corresponding $20\,\mathrm{MHz}$ derived from the same experiment recording, while the $20\,\mathrm{MHz}$ input retains only the central subcarriers.

To illustrate the characteristics of the processed \ac{CSI}, Fig.~\ref{fig:amp_phase_activity} shows the channel response amplitude and phase over frequency and over the full activity sequence in experiment $\beta_1$. 
As seen in the figure, both components exhibit clear temporal variations across different activities.
The horizontal axis represents time in seconds, while the vertical axis represents the subcarrier index.
The dashed lines mark the transitions between consecutive activities, and the corresponding activity labels are shown at the top of each plot.
In Fig.~\ref{fig:amp}, the amplitude presents activity-dependent temporal variations across the subcarriers.
Although the overall amplitude remains relatively smooth, clear vertical patterns appear during several activity intervals.
In particular, the longer falling interval produces a visibly different response from shorter activities such as shaking, rotating, and jumping.
Fig.~\ref{fig:phase} shows the phase after preprocessing and temporal smoothing.
The processed phase also contains activity-dependent temporal structures, although a small number of samples still show residual phase outliers close to $-180^\circ$ or $180^\circ$.

% Using these preprocessed inputs, we first evaluate DT-Pose under different input representations and bandwidth settings.
Using these preprocessed inputs, we evaluate DT-Pose for the two independent 20\,MHz recordings and compare the amplitude-only and joint amplitude–phase input representations.
We then analyze the results by activity and keypoint, examine the training behaviour, and finally present representative qualitative pose estimates.
%During evaluation, we use \ac{PCK} to measure the percentage of correctly localized keypoints under different distance thresholds.

\subsection{Training and Validation Performance}
\label{subsec:quantitative_results}
\begin{figure}[t]
\centering
\subfloat[Training and validation loss.]{
    \includegraphics[width=0.41\linewidth]{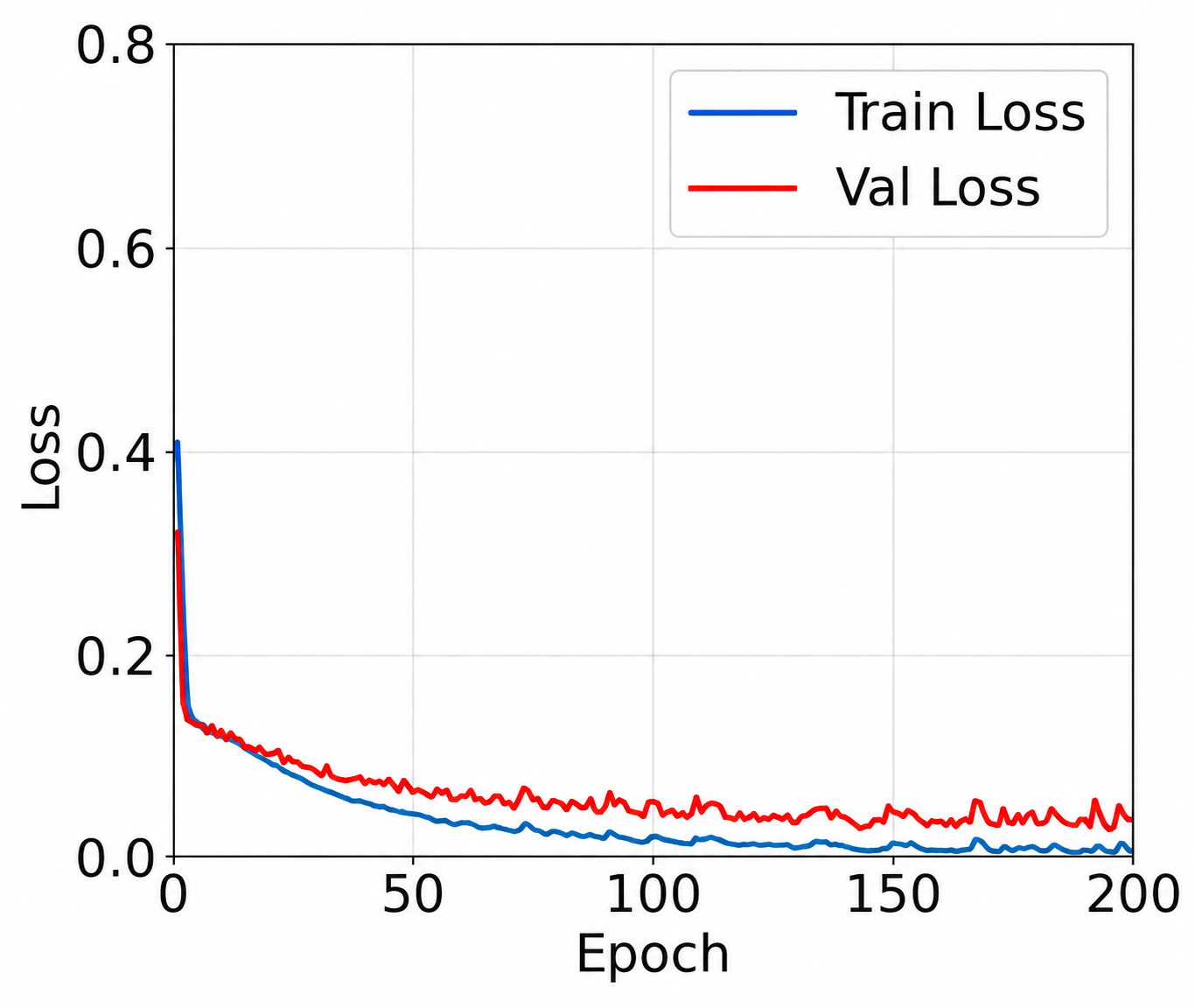}
    \label{fig:trainval}
}
\subfloat[Validation PCK curves.]{
    \includegraphics[width=0.56\linewidth]{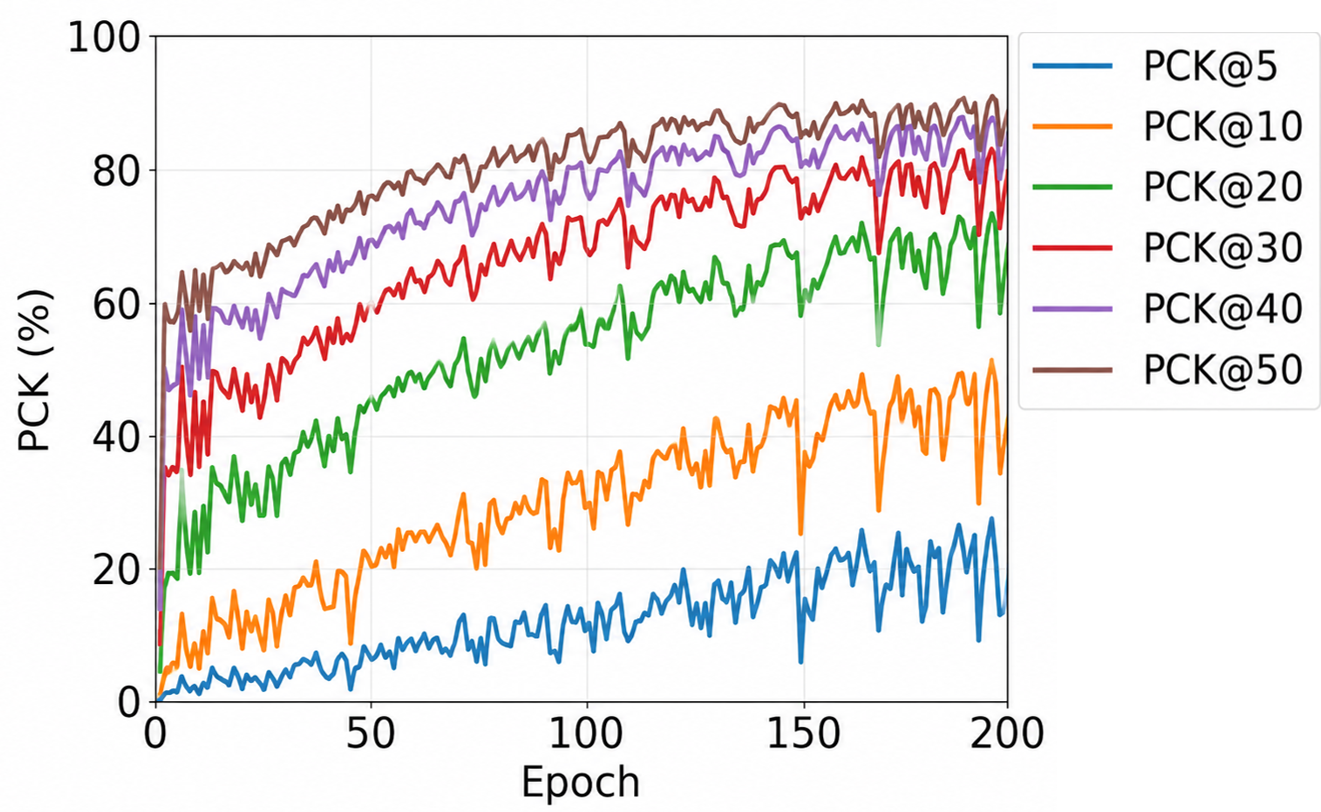}
    \label{fig:pckcurves}
}
\caption{
%\textcolor{orange}{new figure of using non-preprocessing}
Training and validation performance in terms of loss and \ac{PCK} for joint amplitude--phase training.
The validation loss is used to select the final model checkpoint.
}
\vspace{-7pt}
\label{fig:training_pck}
\end{figure}

The learning behaviour of the joint amplitude--phase model is shown in Fig.~\ref{fig:training_pck}. 
Over the epochs, the training and validation losses gradually decrease, while the validation \ac{PCK} values improve and eventually become stable.
The curves show no clear late-stage divergence between training and validation performance, suggesting that the model continues to learn useful pose-related features. Most of the improvement occurs before approximately epoch $180$. %, after which the curves change only slightly. 
After training, the metrics are listed in Table~\ref{tab:input_bandwidth_results} together with the number of available samples after interpolation.

% \textcolor{orange}{\textbf{Reviewer~2 also suggested phase-preprocessing ablation}}

For the experiment $\beta_1$, the results show that \ac{lpr} does not consistently improve pose-estimation performance. Without \ac{sg} filtering, retaining the linear phase component improves the joint amplitude--phase \ac{PCK} from $41.0\%$, $76.3\%$, and $86.2\%$ to $51.5\%$, $83.2\%$, and $91.2\%$ at PCK@10, PCK@30, and PCK@50, respectively. This suggests that the fitted linear phase component may contain useful propagation-related information in addition to synchronization-induced distortions. 
%The reason of different \ac{PCK} of using phase or not on the training is that the best-validation checkpoint occurred at different epochs.  
In contrast, when \ac{sg} filtering is applied, \ac{lpr} improves PCK@10 from $44.5\%$ to $50.0\%$, PCK@30 from $77.5\%$ to $83.2\%$ and PCK@50 from $87.4\%$ to $90.7\%$, indicating that an interaction between temporal smoothing and phase detrending might help. These results highlight that the \ac{sg} filter and \ac{lpr} preprocessing together do not provide any benefit compared to \ac{DL} applied directly to raw amplitude and phase data.
%\textcolor{blue}{
However, combining \ac{sg} filter and \ac{lpr} improves the joint amplitude-phase performance compared with either operation used individually.
This indicates that the combined \ac{sg}+\ac{lpr} configuration remains beneficial relative to the single-operation alternatives.
%}
%In experiment $\beta_2$, although only $4\,072$ samples are available, joint amplitude--phase training improves \ac{PCK}@10, \ac{PCK}@30, and \ac{PCK}@50 from $19.8\%$, $54.3\%$, and $70.0\%$ to $30.8\%$, $69.5\%$, and $74.6\%$, respectively.
Finally, for the $\beta_2$ experiment, joint amplitude--phase training consistently
outperforms amplitude-only training despite the smaller dataset.
%, comparing the effects of input representation and signal bandwidth together with the number of available samples after interpolation.

% \begin{table}[t]
% \centering
% \caption{
% Comparison of \ac{PCK} of different experiments.  Each entry is reported as joint amplitude--phase training / \textit{\scriptsize{AMPLITUDE-ONLY TRAINING}}. 
% The best reported value in each column is in bold.
% }
% \label{tab:input_bandwidth_results}
% \setlength{\tabcolsep}{5pt}
% \renewcommand{\arraystretch}{1.00}
% \begin{tabular}{llcccc}
% \toprule
% Exp. & Bandwidth & Samples
% & PCK@10 $\uparrow$
% & PCK@30 $\uparrow$
% & PCK@50 $\uparrow$ \\
% \midrule
% $\beta_1$ & $20\,\mathrm{MHz}$
% & 12\,677
% & \textbf{50.0} / \textit{48.2}
% & \textbf{83.2} / \textit{83.1}
% & \textbf{90.7} / \textit{90.7} \\
% \midrule
% $\beta_2$ & $20\,\mathrm{MHz}$
% & 4\,072
% & {30.8} / \textit{19.8}
% & {69.5} / \textit{54.3}
% & {74.6} / \textit{70.0} \\
% \bottomrule
% \end{tabular}
% \end{table}

\begin{table}[t]
\centering
\caption{
Comparison of \ac{PCK} under different CSI preprocessing configurations
and experimental recordings. \\
Each entry is reported as joint amplitude--phase training /
\textit{\scriptsize{AMPLITUDE-ONLY TRAINING}}.
}
\label{tab:input_bandwidth_results}
\setlength{\tabcolsep}{3.2pt}
\renewcommand{\arraystretch}{1.00}

\begin{tabular}{llcccc}
\toprule
Exp. & Preprocessing & Samples
& PCK@10 $\uparrow$
& PCK@30 $\uparrow$
& PCK@50 $\uparrow$ \\
\midrule

$\beta_1$
& None
& 12\,677
& \textbf{51.5} / \textit{44.1}
& \textbf{83.2} / \textit{80.8}
& \textbf{91.2} / \textit{90.1} \\

$\beta_1$
& LPR
& 12\,677
& 41.0 / \textit{42.4}
& 76.3 / \textit{79.6}
& 86.2 / \textit{88.8} \\

$\beta_1$
& SG
& 12\,677
& 44.5 / \textit{39.1}
& 77.5 / \textit{78.7}
& 87.4 / \textit{88.5} \\

$\beta_1$
& SG+LPR
& 12\,677
& 50.0 / \textit{48.2}
& \textbf{83.2} / \textit{83.1}
& 90.7 / \textit{90.7} \\

\midrule

$\beta_2$
& SG+LPR
& 4\,072
& 30.8 / \textit{19.8}
& 69.5 / \textit{54.3}
& 74.6 / \textit{70.0} \\

\bottomrule
\end{tabular}
\end{table}

\begin{table}[t]
\centering
\caption{
Per-activity performance assessed as in \eqref{eq:pck} for the $20\,\mathrm{MHz}$ joint amplitude--phase model. 
\ac{PCK} values are reported in percentage, and higher values are better.
}
\label{tab:activity_results_20mhz_joint}
\renewcommand{\arraystretch}{1.08}
\begin{tabular}{llccc}
\toprule
Number & Activity & PCK@10 $\uparrow$ & PCK@30 $\uparrow$ & PCK@50 $\uparrow$ \\
\midrule

1 & Moving         & 84.15 & 97.28 & 99.56 \\

2 & Shaking    & 60.24 & 87.33 & 93.44 \\

3 & Bending        & 68.48 & 93.05 & 97.24 \\

4 & Stretching     & 59.35 & 88.38 & 94.30 \\

5 & Rotating       & 37.23 & 72.76 & 82.11 \\

6 & Crouching      & 50.13 & 84.37 & 93.26 \\

7 & Jumping        & 30.19 & 69.72 & 84.30 \\

8 & Falling        & 30.30 & 72.92 & 85.48 \\

\bottomrule
\end{tabular}
\end{table}

% Table~\ref{tab:pck_per_keypoint} reports the performance for each COCO17 keypoint. The lower-body joints generally achieve higher accuracy than the upper-limb endpoints. In particular, the knees, hips, and ankles obtain PCK@50 values of approximately $89\%$--$92\%$. The right knee achieves the highest PCK@10 value of $42.94\%$, while the left knee reaches the highest PCK@50 value of $91.70\%$.
% The elbows and wrists are the most difficult keypoints to estimate. The left and right wrists obtain PCK@10 values of only $15.78\%$ and $18.85\%$, respectively. This is reasonable because distal upper-limb joints often move rapidly over a large spatial range and may produce weaker or less stable changes in the radio channel. Their lower accuracy may also be related to the limited number and diversity of training samples.

% \input{Tables/table_metrics3}

\subsection{Pose Results}
\label{subsec:qualitative_results}
% \textcolor{red}{excellent section}
% \textcolor{orange}{Thank you this comment made me happy one night long!}

\begin{figure}[t]
\centering
% \hspace{-11pt}
\subfloat[Successful examples]{
    \includegraphics[width=0.48\linewidth]{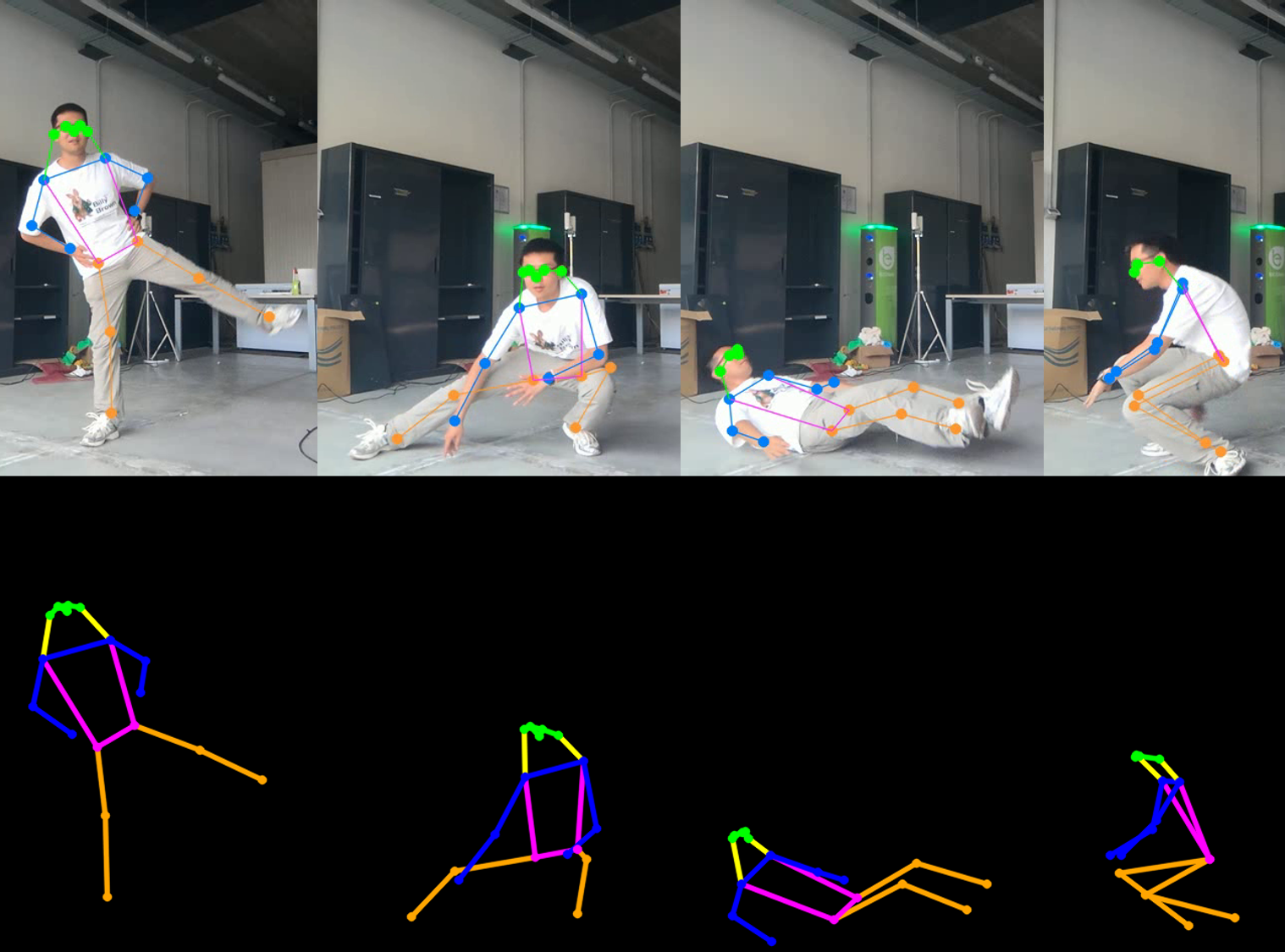}
    \label{fig:poses_good}
}
\subfloat[Failure examples]{
    \includegraphics[width=0.48\linewidth]{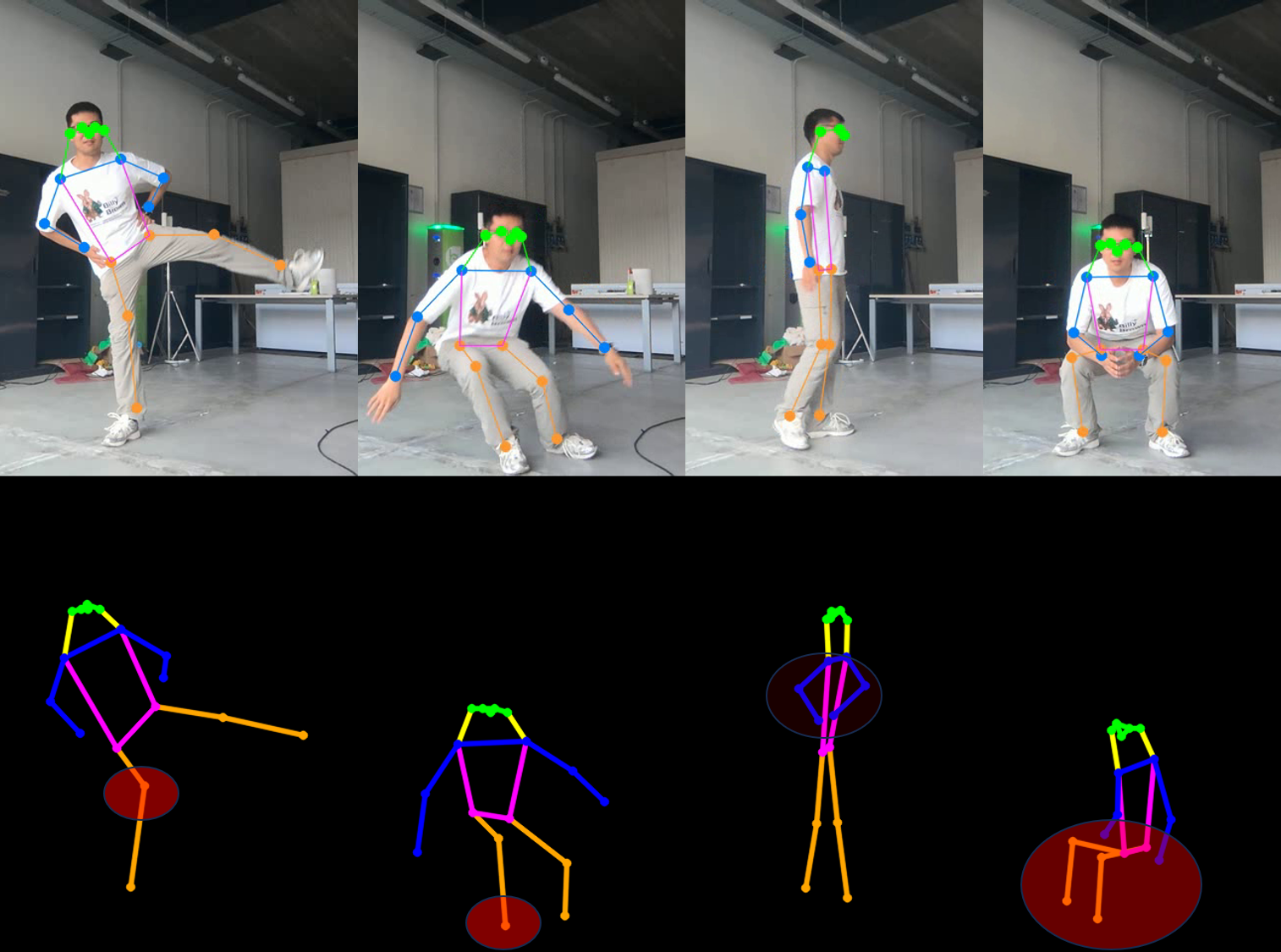}
    \label{fig:poses_bad}
}
\caption{
Representative qualitative pose estimation results.
For each example, the first row shows the camera image with the \ac{GT} skeleton, while the second row shows the pose predicted by the \ac{CSI}-based pipeline.
The red regions highlight joints with large prediction errors.
}
\label{fig:poses}
\end{figure}

Table~\ref{tab:activity_results_20mhz_joint} reports the per-activity performance of the $20\,\mathrm{MHz}$ joint amplitude--phase model. The model performs best for bending and moving, while rotating, jumping, and falling are more difficult. These activities involve larger body motion, rapid transitions, or more complex limb configurations, which can increase keypoint localization errors.
Furthermore, ten representative pose-estimation examples are illustrated in Fig.~\ref{fig:poses}, highlighting examples of successful  (Fig.~\ref{fig:poses_good}) and imperfect (Fig.~\ref{fig:poses_bad}) estimates.
For each example, the first row shows the camera image with the \ac{GT} skeleton, while the second row shows the corresponding pose predicted from the \ac{prs}-based \ac{CSI}.
The successful examples in Fig.~\ref{fig:poses_good} show that the proposed pipeline can recover a range of body configurations, including standing, bending, crouching, falling, and arm-raising poses.
In most cases, the predicted skeleton follows the main body orientation and preserves the relative positions of the torso and limbs.
This indicates that the \ac{CSI} input contains useful spatial information about human motion and posture.

The failure cases in Fig.~\ref{fig:poses_bad} mainly occur at limb endpoints, especially the wrists and ankles.
In the first and second failure cases, there is one joint having a poor alignment to the \ac{GT}.
For example, in the third case, the subject is rotating, but the predicted skeleton faces the wrong direction.
For this type of activity, which includes rotation, the depth dimension plays a significant role. Additional spatial diversity, for example through multiple gNBs or receive antennas, may help resolve such orientation ambiguities.
%3D sensing with additional devices distributed uniformly around the user is likely required to improve the geometric factor.
A similar orientation error appears in the fourth crouching example.
% In the fifth failure example, the subject is jumping, but the predicted skeleton places one arm in an incorrect raised position.
Overall, the model estimates the main body structure well, while fine-grained limb localization and orientation remain challenging in the 2D domain with a small number of dataset samples.

\section{Conclusion}
\label{sec:conclusion}

This work investigated whether \ac{CSI} extracted from standardized \ac{5g} \ac{prs} can support device-free 2D human pose estimation.  
At $20\,\mathrm{MHz}$, joint amplitude--phase training improves strict keypoint localization compared with amplitude-only training, with the largest gain observed for \ac{PCK}@10.  
The pre-processing comparison further showed that dedicated \ac{sg} filter and \ac{lpr} are not required with the best performance obtained by directly using the amplitude and phase of the \ac{LS}-estimated \ac{CSI}.
The proposed pipeline also recovers the main body structure across different activities.
% with the strongest performance obtained for the hips, knees, and ankles.
% The per-keypoint and qualitative results also reveal the main limitations of the current system. 
The wrists and elbows are more difficult to localize because they often move rapidly and cover a larger spatial range.  
Some failure cases also contain errors in body orientation or hand positions. 

%\textcolor{orange}{
%\sout{
%In addition, the preliminary $40\,\mathrm{MHz}$ dataset contains fewer samples and fewer activities than the $20\,\mathrm{MHz}$ dataset. 
%Therefore, the current bandwidth results should be interpreted as an initial study rather than a fully controlled comparison.}
%}

% Future work will consider balanced datasets with the same subjects, activities, and different bandwidths. 
Future work will investigate generalization across multiple subjects, environments, activity sequences, and radio geometries, using larger and different bandwidths.
%The experiment with a bandwidth at $40\,\mathrm{MHz}$ is currently under investigation. 
More robust phase processing, longer training schedules and static pose recordings may further improve the localization of rapidly moving joints. Additional \acp{gnb} or receive antennas could also provide greater spatial diversity.

% \begin{figure}[t]
% \centering
% \subfloat[AMP.]{
%     \includegraphics[width=0.95\linewidth]{FigPdfPng/amp_40MHz.png}
%     \label{fig:poses_good}
% }
% \\
% \subfloat[PHASE.]{
%     \includegraphics[width=0.95\linewidth]{FigPdfPng/phase_40MHz.png}
%     \label{fig:poses_bad}
% }
% \caption{40 MHz \textcolor{orange}{Testing if there is artifact or noise at the end of the subcarriers (in the phase...) But I cannot really tell. This figure is temporary and to be removed after review}}
% \label{fig:40MHz}
% \end{figure}

\section*{Acknowledgment}
% The authors acknowledge the use of AI-assisted tool  (ChatGPT) as a generative language assistant, for translation of Chinese drafts into English, for grammar correction and refinement of English sentence formatting. All generated content was reviewed and revised by the authors to ensure scientific accuracy and consistency. 
The authors acknowledge the use of an AI-assisted tool (ChatGPT) for English-language proofreading. All AI-generated content was carefully reviewed and revised by the authors to ensure scientific accuracy and consistency.
% \textcolor{red}{explain exactly how you used it}
% \textcolor{orange}{I translated lots of chinese keypoints of my notes into English, also prompting it with "easy academic english", "also after this refining my expression, output in latex format", and also ask him to tell me what is missing in my description. After translating, It will suggest me how to write. I copy the text, and then edit the text. The abstract section is created with prompt "Cover the keypoints and write me a scientific abstract, while the old abstract was written like this......"}

\bibliographystyle{IEEEtran}
\bibliography{refs}

\end{document}